\documentclass{emulateapj}

\def\Meszaros{M\'esz\'aros~}

\def\simg{\mathrel{\rlap{\raise 0.511ex \hbox{$>$}}{\lower 0.511ex \hbox{$\sim$}}}}
\def\siml{\mathrel{\rlap{\raise 0.511ex \hbox{$<$}}{\lower 0.511ex \hbox{$\sim$}}}}

\def\calN{{\cal N}}      

\def\bLE{\beta_{LE}}     
\def\bog{\beta_{o\gamma}}

\def\tad{t_{ad}}         
\def\tsy{t_{sy}}         
\def\tsyi{t_{sy,i}}      
\def\tic{t_{ic}}         
\def\tici{t_{ic,i}}      
\def\tci{t_{rad}}        
\def\trad{t_{rad}}       

\def\tB{t_B}             
\def\ti{t_I}             
\def\to{t_o}             
\def\tang{t_{ang}}       

\def\gi{\gamma_i}        
\def\go{\gamma_o}        
\def\gc{\gamma_{cr}}     
\def\gm{\gamma_m}        
\def\gp{\gamma_p}        
\def\gx{\gamma_X}        
\def\gad{\gamma_{ad}}    
\def\gsy{\gamma_{sy}}    

\def\dtg{\delta t_\gamma}
\def\dteps{\delta t_ \eps} 

\def\Ep{E_\gamma}        
\def\E5{E_{\gamma,5}}    
\def\Fp{F_p}             
\def\eps{\epsilon}       
\def\esy{\epsilon_{sy}}  
\def\fe{f_\epsilon}      
\def\Fe{{\cal F}_\epsilon} 
\def\em{\varepsilon_m}   
\def\fm{f_m}             
\def\ep{\varepsilon_p}   
\def\fp{f_p}             

\def\te{t_{\gamma \epsilon}}   
\def\tp{\hat{t}_{\gamma \epsilon}}  
\def\tesy{\te^{(sy)}}    
\def\tead{\te^{(ad)}}    
\def\teic{\te^{(ic)}}    
\def\tgx{t_{\gamma-10k}} 
\def\tpx{\hat{t}_{\gamma-10k}}  
\def\tgxsy{\tgx^{(sy)}} 
\def\tgxad{\tgx^{(ad)}} 
\def\tpxad{\tpx^{(ad)}} 
\def\tgxic{\tgx^{(ic)}} 
\def\tt{\tilde{t}}       
\def\ttic{t_{2/3 \rightarrow 1}} 
\def\tcr{t_{cr}}         

\def\tpk{t_p}            

\def\epstoEp{\left(\h \displaystyle \frac{\eps}{\Ep} \h \right)}

\def\ds{\displaystyle}
\def\h{\hspace*{-1mm}} \def\hh{\hspace*{-2mm}} \def\hhh{\hspace*{-3mm}}
\def\hhhh{\hspace*{-4mm}} \def\hhhhh{\hspace*{-5mm}}
\def\ra{\longrightarrow}
\def\ttimes{\h\times\h}

\begin{document}

\parskip 5pt
\topmargin 10mm

\title{The Synchrotron Low-Energy Spectrum Arising from the Cooling of Electrons in Gamma-Ray Bursts}

\author{A.D. Panaitescu and W.T. Vestrand}

\affil{Intelligence and Space Research, Los Alamos National Laboratory, Los Alamos, NM 87545, USA}

\vspace{2mm}
\begin{abstract}

 This work is a continuation of a previous effort (Panaitescu 2019) to study the cooling of relativistic electrons through 
 radiation (synchrotron and self-Compton) emission and adiabatic losses, with application to the spectra and light-curves
 of the synchrotron Gamma-Ray Burst produced by such cooling electrons.
 Here, we derive the low-energy slope $\bLE$ of GRB pulse-{\sl integrated} spectrum and quantify the implications of the
 measured distribution of $\bLE$.

 If the magnetic field lives longer than it takes the cooling GRB electrons to radiate below 1-10 keV, then radiative 
 cooling processes of power $P(\gamma) \sim \gamma^n$ with $n \geq 2$ ($\gamma$ is the electron energy), 
 i.e. synchrotron and inverse-Compton (iC) through Thomson scatterings, lead to a soft low-energy spectral slope 
 $\bLE \leq -1/2$ of the GRB pulse-integrated spectrum $\Fe \sim \eps^{\bLE}$ below the peak-energy $\Ep$,
 irrespective of the duration of electron injection $\ti$. 
 iC-cooling dominated by scatterings at the Thomson--Klein-Nishina transition of synchrotron photons below $\Ep$
 has an index $n = 2/3 \rightarrow 1$ and yield harder integrated spectra with $\bLE \in [0,1/6]$, while adiabatic 
 electron-cooling leads to a soft slope $\bLE = -3/4$. 

 {\sl Radiative} processes that produce soft integrated spectra can accommodate the harder slopes measured by CGRO/BATSE and 
 Fermi/GBM {\sl only if} the magnetic field life-time $\tB$ is shorter than the time during which the typical GRB electron 
 cools to radiate below 10 keV (i.e. less than several radiative cooling timescales $\trad$ of that typical electron). 
 In this case, there is a one-to-one correspondence between $\tB$ and $\bLE$.
 To account for low-energy slopes $\bLE > -3/4$, {\sl adiabatic} electron-cooling requires a similar restriction on $\tB$. 
 In this case, the diversity of slopes arises mostly from how the electron-injection rate varies with time 
 (temporal power-law injection rates yield power-law low-energy GRB spectra) and not from the magnetic field timescale.

\end{abstract}

\vspace{3mm}
\section{\bf Introduction}



\vspace{2mm}
\subsection{\bf GRB-Pulse Temporal Properties}

 GRB observations (e.g. Fenimore et al 1995, Norris et al 1996, Lee et al 2000) have established some essential/basic 
features of GRB pulses: \\
$i)$ they peak earlier at higher energies, \\
$ii)$ they are time-asymmetric, rising faster than they fall, with a rise-to-fall time ratio $t_r/t_f$ in the range $(0.1-0.9)$,\\
$iii)$ their temporal asymmetry is {\sl on average} energy-independent, and \\
$iv)$ they last longer at lower energies, having a pulse duration--energy dependence $\dtg \sim \eps^{-0.4}$. 


 Figures 5 and 6 of Panaitescu 2019 (P19) provide a limited assessment of the ability of adiabatic and synchrotron electron 
cooling to account for the above pulse features: \\
 $i)$ peaks occurring earlier at higher energies is a trivial consequence for any electron cooling process, \\ 
 $ii)$ both cooling processes yield pulses that are more time-symmetric at higher energies, in conflict with observations
     of most GRB pulses, \\ 
 $iii)$ if the pulse duration $\dtg$ dependence on energy arises from only electron cooling then, for a {\sl constant
   magnetic field}, adiabatic cooling yields $\dtg \sim \eps^{-0.4}$ (weaker than expected analytically) and synchrotron
   cooling leads to $\dtg \sim \eps^{-0.5}$ (as expected), both being compatible with GRB observations.

 The {\sl geometrical curvature} of the emitting surface leads to a spread in emission angles over the spherical surface  
of the GRB ejecta, increases all observer-frame timescales by $\sim 50$\%. Additionally, it delays the arrival-time of a photon 
emitted (toward the observer) from the fluid moving at a larger angle (relative to its radial direction of motion) 
due to a longer path to observer and reduces its energy (due to a lower relativistic boost). Therefore, the integration 
of emission over the angle of the fluid motion softens continuously the received emission by delaying the arrival of photons 
of lesser energy. 
 
 Numerical calculations (P19) of GRB pulses show that the angular integration associated with the geometrical curvature 
of the emitting surface has the following effects on the pulse properties:  \\ 
 $i)$ contributes to pulses peaking earlier at higher energies (which is the continuous emission softening described above), \\
 $ii)$ mitigates the wrong trend of pulses to be more time-symmetric at higher energy when synchrotron-cooling is dominant 
   because, in that case, the synchrotron-cooling timescale $\tsy$, being shorter than the adiabatic-cooling timescale 
   $\tad = 3\,\tang$ (see \S\ref{pulsefall}), is also (likely) smaller than the angular time-spread $\tang$, 
   thus the pulse rise and fall timescales 
   $t_r$ and $t_f$ are set by the angular integration, which does not induce an energy dependence of the ratio $t_r/t_f$, \\
 $iii)$ is unable to compensate for pulses being more time-symmetric at higher energy when adiabatic cooling is dominant
   because the angular time-spread $\tang$ is smaller than the adiabatic-cooling timescale $\tad$, thus the pulse-rise and 
   fall timescales are not changed much by the angular integration, \\
 $iv)$ leads to pulses lasting longer at lower energies (owing to the progressive softening of the received emission) and
   induces a pulse-duration energy-dependence $\dtg(\eps) \sim \eps^{-0.4}$ that is similar to that produced by each
   cooling process for a constant magnetic field. 

 The integration of the received emission over the equal photon-arrival time is effective only if the emitting region 
extends an angle larger than $\Gamma^{-1}$, the inverse of the Lorentz factor at which that region moves toward the observer, 
and its effect is diminished if the emitting region is a {\sl bright-spot} of angular extent well below $\Gamma^{-1}$.
 Therefore, the above evaluation of the pulse properties resulting when electron cooling is synchrotron-dominated applies 
only to GRB pulses that arise from bright-spots. However, given that the angular integration has little effect on the
pulse properties when the electron cooling is adiabatic, the previous evaluation of those pulse properties is correct
for both a bright-spot and an uniformly-bright surface.

 Consequently, if the trend of numerically-calculated pulses to be more symmetric at higher energies is firmly established 
then its incompatibility with observations (for either electron cooling process) favors the hypothesis that GRB pulses
arise from a uniformly-bright surface and that the electron cooling is synchrotron-dominated, i.e. disfavors a bright-spot 
origin for GRB pulses and an adiabatic-dominated electron cooling.

 However, the pulse timescales and properties depend on the evolution of the electron injection-rate $R_i$ and of the magnetic 
field $B$ (the effect of monotonically-varying such quantities is illustrated by the pulse shapes and durations shown in 
figures 5 and 6 of P19), thus, a comprehensive numerical study of the pulse properties expected for various electron cooling 
processes might (not guaranteed) identify evolving injection-rates $R_i(t)$ and magnetic fields $B(t)$ that accommodate all 
the basic GRB pulse features.

 This work shows the effect of a power-law evolving injection rate $R_i(t) \sim t^y$ on the GRB pulse-{\sl integrated}
spectrum, with emphasis on the diverse low-energy slopes that can be obtained from a decreasing $R_i$ in the case of adiabatic 
electron cooling. A decreasing magnetic field $B(t)$ is important for reconciling with observations the pulse-duration 
dependence on energy resulting when the electron cooling is dominated by scatterings at the Thomson--Klein-Nishina transition
of the synchrotron photons below the peak-energy $\Ep$ of the GRB spectrum.

\vspace{2mm}
\subsection{\bf GRB Low-Energy Spectrum}

 The GRB low-energy slope $\bLE$ (of the energy spectrum below its peak-energy $\Ep$) is measured by fitting the GRB count 
spectrum with various emipirical functions: \\
$i)$ a pure power-law (PL), \\
$ii)$ a power-law with an exponential cut-off (CPL), which is the Band function with a large high-energy spectral slope, \\
$iii)$ the Band function, which is a broken power-law with a fixed width for the transition between the asymptotic power-laws, \\
$iv)$ a smoothly broken power-law (SBPL), which has a free parameter for the width of the transition between the low- and 
   high-energy power-laws.

\vspace{1mm}
\subsubsection{Power-Law GRB Low-Energy Spectrum}

 Preece at al (2000) have analyzed 5500 pulse-integrated spectra at 25 keV -- 2 MeV of the 156 brightest (in peak flux or fluence) 
CGRO/BATSE GRBs, with 80\% of bursts being fit with the Band and the SBPL functions, and have found a distribution for the 
low-energy slope of the pulse-peak spectra that is approximately a Gaussian 
\begin{equation}
\hh  \left[F_\eps (\eps < \Ep) \sim \eps^{\bLE}\right] \; P(\bLE) \sim \exp\left\{-\frac{(\bLE-\beta_o)^2}{2\sigma^2}\right\}
\label{PbLE}
\end{equation}
peaking at $\beta_o = 0.0$ and with a dispersion $\sigma \simeq 0.40$ (half-width at half-maximum of 0.45). 

 For a larger sample of 8093 time-resolved spectra from 350 bright BATSE bursts fit with the CPL, Band, and SBPL, Kaneko et al 
(2006) found a distribution of the low-energy slope $P(\bLE)$ (for their GOOD sample) similar to that of Preece et al (2000), 
with a weighted mean\footnotemark $\beta_o = 0.00$ and a variance $\sigma = 0.14$. A minority of 366 time-resolved 
spectra were fit with a PL and are significantly softer, with $\beta_o^{(soft)} = -0.74 \pm 0.19$.
\footnotetext{This is the variance-weighted average of the three {\sl median} slopes found for the above three fitting functions.
However, the individual distributions do not display any visible skewness, thus the median slope should be very close to the 
variance-weighted average slope, for each of the three sets.}
 
 The "parameter error" criterion used by Poolakkil et al (2021) for selecting the fitting function for Fermi/GBM peak-flux 
spectra at 10 keV--1 MeV leads to a bimodal distribution for the low-energy spectral slope $\bLE$ (of their GOOD sample):
$i)$ PL fits were used for the peak-flux spectra of 2287 bursts, leading to a median slope $\beta_o^{(soft)} = -0.50 \pm 0.18$,
$ii)$ CPL, Band, and SBPL functions were used to fit the 1.0 s peak-flux spectra of 1897 bursts, leading to a median 
  spectral slope $\beta_o^{(hard)} = 0.31 \pm 0.17$.

 The analyses of Kaneko et al (2006) and Poolakkil et al (2021) are similar, as they used the same fitting functions and
retained only those fits that led to lower parameter errors (the GOOD sample) and which had a higher statistical significance 
(the BEST sample), yet the two distributions of low-energy indices $\bLE$ are incompatible with each other, with the BATSE 
bursts being softer than the non-PL GBM bursts. The same is true for the sample of softer bursts that were fit with a PL.

 Poolakkil et al (2021) attribute the bimodality of the $P(\bLE)$ distribution to the PL model being sufficient for the spectral 
fitting of the lower fluence GBM peak-flux spectra, probably because the break to a softer spectrum above the peak-energy $\Ep$ 
is lost for low S/N measurements at higher energies, which leads to a softer best-fit spectrum over the entire GBM window. 


 Given that the bimodality of the $P(\bLE)$ distribution for GBM bursts is "compromised" by the "insensitivity" of PL fitting
to the true hardness of low-energy spectra for dimmer bursts, we will make further use of the $P(\bLE)$ distribution for BATSE 
bursts, and we will forget (and forgive) the excitement caused by that the peaks of the GBM bimodal distribution at slopes
0.31 and -0.50 are very close to or exactly at the values expected for synchrotron emission from uncooled and cooled electrons, 
respectively.

\vspace{1mm}
\subsubsection{Broken Power-Law GRB Low-Energy Spectrum}

{\it 
 Before proceeding, we should note that strong evidence for electron cooling in the GRB low-energy spectra has been found by
fitting the fluence-brightest GRBs spectra below the peak-energy $\Ep$ with a broken power-law instead of a single power-law : \\
$i)$ For 14 bright Swift GRBs with simultaneous observations by XRT (0.3--10 keV) and BAT (10--150 keV),
  Oganesyan et al (2017) have found that 2/3 of 86 instantaneous spectra are better fit with a double SBPL 
  (three power-law segments) having a lower-energy break $E_b \in (2,8)$ keV (and a peak-energy $\Ep \in (30,500)$ keV, thus
   $E_b \simeq 0.03 \;\Ep$) and spectral indices $\beta_o^{(low)} = 0.33 \pm 0.35$ and $\beta_o^{(high)} = -1.46 \pm 0.20$ 
   below and above $E_b$, respectively, with most of the remaining spectra fit adequately by a CPL with an average low-energy
   slope $\beta_o = -0.08 \pm 0.23$,  \\ 
$ii)$ For ten Fermi (10 keV--3 MeV) long GRBs with the largest fluence, Ravasio et al (2019) have found that 70\% of 75 instantaneous 
  spectra are better fit by a double SBPL with a break-energy $E_b \in (20,400)$ keV (and peak-energy $\Ep \in (300,3000)$ keV, 
  thus $E_b \simeq 0.1 \;\Ep$) and spectral indices $\beta_o^{(low)} = 0.42 \pm 0.16$ and $\beta_o^{(high)} = -0.52 \pm 0.20$,
  while the remaining 30\% of spectra are well-fit by a SBPL with an average low-energy slope $\beta_o = -0.02 \pm 0.19$. 

 Both of these works present evidence for a cooling-break of the synchrotron soectrum at energy $E_b < \Ep$ corresponding to the 
ynchrotron emission from the lowest-energy cooled electrons, with the spectral indices below and above $E_b$ being very close to 
the expectations for the emission from synchrotron-cooling electrons: $\beta_o^{(low)} = 1/3$ and $\beta_o^{(high)} = -1/2$.

 The spectra simulated by Toffano et al (2021) have shown that 
$i)$ such cooling breaks require SBPL fits if the burst is sufficiently fluence-bright ($\Phi = 3 \times 10^{-4} \, {\rm erg/cm^2}$),
$ii)$ for average or dim bursts ($\Phi \leq 3 \times 10^{-5} \, {\rm erg/cm^2}$), the Band function provides a good fit because 
of the low S/N ratio, and 
$iii)$ a Band fit yields intermediate low-energy spectral slopes, transiting from $\bLE = -1/2$ (for the emission from the 
synchrotron-cooling tail) to $\bLE = 1/3$ (for the emission from uncooled electrons) when the cooling energy $E_b$ is increased 
from $0.01\, \Ep$ to $0.1\,\Ep$, which would explain the diversity of slopes measured by BATSE and Fermi (Equation \ref{PbLE}). 
However, this interpretation requires that the cooling energy of all bursts satisfies $E_b \in (0.01,0.1)\, \Ep$ because, otherwise, 
a low break-energy $E_b < 0.01\, \Ep$ would yield a peak of the $P(\bLE)$ distribution at $\bLE = -1/2$, while a high break-energy
$E_b \in (0.1,1)\, \Ep$ would lead to a peak at $\bLE = 1/3$, none of which is seen.

}

\vspace{1mm}
\subsubsection{What is done here}

 In this work, we use the compatibility of the calculated pulse-integrated low-energy spectrum slope and observations (Equation 
\ref{PbLE}) to set upper limits on the life-time $\tB$ of the magnetic field, when the electron-cooling stops (if it is radiative) 
and when the production of synchrotron emission ends, keeping in mind that values of the cooling energy $E_b$ within 
two decades below the peak-energy $\Ep$ could account for the diversity of GRB low-energy slopes. This incomplete complete electron 
cooling was first proposed by Oganesyan et al (2017) and Ravasio et al (2019) as the origin for the observed $P(\bLE)$ distribution.

 The following work builds on that of P19, who have presented an analytical derivation of (and numerical results for) the 
low-energy slope of the {\sl instantaneous} synchrotron spectrum for adiabatic, synchrotron, and inverse-Compton dominated 
electron cooling. Here, we present (for all three electron cooling processes) analytical derivations of the pulse light-curve 
at energies below the GRB's and of the low-energy slope of the pulse-{\sl integrated} synchrotron spectrum.


\vspace{2mm}
\subsection{\bf Limitations of the Standard Synchrotron Model}

\vspace{1mm}
\subsubsection{\sl The Low-Energy Spectral Slope}

 An important shortcoming of the basic synchrotron model for the GRB emission is that it cannot account for low-energy slopes
harder than the $\bLE = 1/3$ displayed by about \\
$i)$ 1/3 of CGRO/BATSE 25 keV--2 MeV time-resolved spectra (Preece et al 2000), \\
$ii)$ 1/10 of the 30 time-integrated 2--20 keV spectra of X-ray Flashes and GRBs observed by BSAX/WFC (Kippen et al 2004) and BATSE, and \\ 
$iii)$ 1/4 of Fermi/GBM peak-flux 10 keV--1 MeV spectra of the BEST sample (Poolakkil et al 2021). 

 Thus, if an yet-unidentified large systematic error $\sigma(\bLE) \simeq 0.3$ does not explain away the low-energy spectral 
slopes harder than $\bLE = 1/3$, then the following formalism for studying the effects of electron cooling on the GRB 
synchrotron emission is relevant for a majority a GRBs but a deviation from that model (or another emission process) is needed
for a substantial fraction of bursts. 

 The shortest departures from that model harden the low-energy slope to $\bLE = 1$ by relying on a very small electron 
pitch-angle $\alpha < \gamma^{-1}$ (with $\gamma$ being the electron Lorentz factor), i.e. a pitch-angle less than the opening 
of the cone into which the cyclotron emission is relativistically beamed, as proposed by Lloyd \& Petrosian (2000), 
or on a very small length-scale for the magnetic field, $\lambda_B < \rho_L/ \gamma$ (with $\rho_L$ being the electron gyration 
radius), so that electrons are deflected by angles less than $\gamma^{-1}$ and produce a "jitter" radiation, as proposed by 
Medvedev (2000). A hard slope $\bLE = 1$ is obtained if the GRB emission is the upscattering of self-absorbed lower-energy 
synchrotron photons (Panaitescu \& \Meszaros 2000), but the $\eps F_\eps$ spectrum of the upscattered emission may be too 
broad compared to real GRB spectra.

 In addition to these models that employ synchrotron emission and explain measured low-energy slopes harder than $\bLE = 1/3$, 
a photospheric black-body component (proposed by e.g. \Meszaros \& Rees 2000, used to account for most of spectrum of GRB 090902B 
by Ryde et al 2010, but being in general a sub-dominant component, e.g. Axelsson et al 2012 for GRB 110721A) can yield low-energy 
spectra as hard as $\bLE = 2$, while a combination of synchrotron and thermal emission can lead to intermediate low-energy slopes 
$\bLE \in (1/3,2)$ if the photospheric plus synchrotron GRB spectrum is fit with just the Band function.
The issue of some measured low-energy slopes being too hard for the synchrotron model may be also alleviated by the addition 
of a power-law component to the Band (strongest component) plus thermal (weakest component) decomposition (e.g. Guiriec et al 2015),
although that has been proven for only a small number of bursts.

\vspace{1mm}
\subsubsection{\sl Deficiency of Our Treatment}

 A limitation of the following treatment of GRB pulses as synchrotron emission from a population of cooling relativistic 
electrons is that the effect of electron cooling on the pulse spectral evolution is calculated assuming that the typical 
energy $\gi$ of the injected electrons is constant during the GRB pulse. Another default assumption (occasionally relaxed) 
is that the magnetic field $B$ is also constant. These assumptions are needed for an easier calculation of electron 
cooling electrons but they imply that the peak-energy $\Ep$ of the $\eps F_\eps$ instantaneous spectrum is constant and 
so will be the peak-energy of the integrated spectrum, if the low-energy slope is harder than $\bLE = -1$. 

 However, measurements of the pulse spectral evolution (e.g. Crider et al 1997, Ghirlanda, Celotti \& Ghisellini 2003) 
show that the peak-energy $\Ep$ decreases monotonically throughout the pulse.

 Consequently, the following description of the spectral evolution due to electron cooling for a constant typical electron 
energy $\gi$ and a constant magnetic field $B$ is representative for real GRBs displaying a decreasing peak-energy $\Ep$ 
only if that decrease of the best-fit $\Ep$ value is the artifact of fitting the curvature below $\Ep$ of real instantaneous 
spectra with an empirical function of free or fixed smoothness for the transition between the two (low and high energy) 
power-laws.

\vspace{2mm}
\subsection{\bf Magnetic Field Life-Time and Duration of Electron Injection}

 The GRB low-energy slope and the GRB pulse duration (as well as the GRB-to-counterpart relative brightness and counterpart
pulse duration) depend on the magnetic field life-time $\tB$ (real or apparent) and the duration over which relativistic electrons 
are injected into the region with magnetic field.

 For first-order Fermi acceleration at relativistic shocks, the duration $\ti$ of particle injection in the down-stream region 
is the sum of the shock life-time $t_{sh}$ (the time it takes the shock to cross the ejecta shell) and the duration it takes 
for a given particle to be accelerated, i.e. the time for it to diffuse (for a magnetic field perpendicular to the shock front) 
or to gyrate (for a magnetic field parallel to the shock surface) many times in the up-stream and down-stream regions and 
undergo multiple shock-crossings. 

 For magnetic fields generated by turbulence or two-stream instability (Medvedev \& Loeb 1999) at relativistic shocks, 
which decay in the down-stream region, the magnetic field intrinsic life-time $\tB$ would be the shock life-time $t_{sh}$. 
However, if the particle injection is impulsive (shorter-lived) relative to the shock life and lasts $\ti < \tB$, 
then the apparent magnetic field life-time $\tB$ that a particle spends in the magnetic field region would be the time 
that it takes a particle to cross the down-stream region where there is a magnetic field.

 The above suggest that the durations $\tB$ and $\ti$ may be correlated if particles are accelerated and if magnetic fields
are produced at relativistic shocks. For generality (i.e. to include other mechanisms that produce magnetic fields and 
relativistic particles, such as magnetic reconnection - Zhang \& Huirong 2011, Granot 2016), we consider that the two parameters 
$\ti$ and $\tB$ are independent.

\begin{table*}
\caption{\small \vspace*{2mm} \hspace*{10mm} Glossary of more frequently used notations}
\vspace*{5mm}
\centerline{
\begin{tabular}{ll|ll}
  \hline \hline
Electron energy \\
\hline
$\gi$     & typical energy of injected electrons   & $\gm$  & lowest energy of cooled electrons \\
$\gp$     & energy of electrons radiating at $\ep$ & $\gc$  & critical electron energy, where $\tsy (\gc) = \tad$ \\
 \hline
Spectral quantities \\
\hline
$\bLE$    & GRB low-energy slope (below $\Ep$) & $\bog$  &  optical-to-gamma effective spectral slope \\
$\Ep$     & peak energy of GRB $\nu F_\nu$ spectrum      & $\Fp$  &  flux at $\Ep$ \\ 
$\eps$    & observing energy                             & $\esy$ & SY characteristic energy \\
$\fe$     & spectral flux density                        & $\Fe$  & pulse-integrated spectral flux density \\
$\ep$     & peak energy of the $F_\nu$ SY spectrum       & $\fp$  & SY flux at $\ep$ \\
$\em$     & SY energy for the $\gm$ electrons            & $\fm$  & SY flux at $\em$ \\ 
 \hline
Electron timescales \\
\hline
$\tad$    & AD cooling timescale                     &  $\trad$ & radiative cooling timescale of $\gi$ electrons \\ 
$\tsy$    & SY cooling timescale                     & $\tic$   & iC cooling timescale \\
$\tsyi$   & SY cooling timescale for $\gi$ electrons & $\tici$  & iC cooling timescale for $\gi$ electrons \\
$\te$     & transit time from GRB $\Ep$ energy to $\eps$ & $\tgx$  & transit time from GRB to mid X-rays (10 keV)  \\
$\tcr$    & epoch when $\tsy(\gm)=\tad$          \\        
 \hline
Other timescales \\
\hline
$\tB$     & magnetic field life-time  & $\ti$    & electron injection duration \\
$\tpk$    & pulse peak epoch          & $\tang$  & angular spread in photon arrival-time \\
$\dtg$    & duration of GRB pulse     & $\dteps$ & pulse duration at energy $\eps$ \\
  \hline \hline
\end{tabular}
}
\label{Notations}
\end{table*}

 If electrons are re-accelerated (Kumar \& McMahon 2008), the magnetic field life-time $\tB$ used here can be seen a surrogate 
for the re-acceleration timescale, as particles are allowed to cool only for that re-acceleration timescale, thus our assumption 
that synchrotron emission stops at $\tB$ does not affect the following results about the GRB low-energy spectral slope
(or the brightness of the prompt counterpart). 
 However, if the GRB pulse duration is set by the magnetic field life-time $\tB$, electron reacceleration on a timescale 
$t_{re-acc} > \tB$ could lead to GRB pulses longer than $\tB$, thus a finite magnetic field life-time is not completely
equivalent to the electron re-acceleration.
 
 {\bf Table \ref{Notations}} lists the most often notations used here.

\vspace{3mm}
\section{\bf The Electron-Cooling Law}

 For any cooling process, {\sl conservation} of particles during their flow in energy can be written as
\begin{equation}
  \frac{\partial \calN}{\partial t} + \frac{\partial}{\partial \gamma} \left( \calN \frac{d\gamma}{dt} \right) = \calN_i
\label{cons}
\end{equation}
with $\calN (\gamma) = dN/d\gamma$ the particle distribution with energy and
\begin{equation}
 \calN_i (\gi < \gamma) \sim \gamma^{-p}
\end{equation}
the distribution of the {\sl injected} electrons, set to zero below a {\sl typical/lowest} electron energy $\gi$, and 
\begin{equation}
 -\frac{d\gamma}{dt} = Q(t) \gamma^n
\end{equation}
is the electron {\sl cooling law} for the corresponding cooling process. For {\bf AD}iabatic cooling, $Q$ is a constant; 
for {\bf SY}nchrotron cooling, $Q \sim B^2$, with $B$ the magnetic field, because the SY power is proportional 
to the energy density of the virtual photons that are upscattered to SY photons; for {\sl self-Compton} cooling, 
$Q \sim B^2 \tau$, with $\tau$ the optical-thickness to electron scattering, because the inverse-Compton cooling power 
is proportional to the energy density of SY photons.

 Above $\gi$, Equation (\ref{cons}) has a broken power-law solution, the {\sl cooled-injected} electron distribution
having a break at the electron energy $\gc$ where the radiative cooling timescale equals the time elapsed since the 
beginning of electron injection (AD-cooling does not yield a "cooling-break" because the AD-cooling timescale
is slightly larger than the system age). Going to higher energies, the exponent of the cooled-injected distribution decreases 
by unity at $\gc$. The cooled-injected electron distribution is of importance for calculating the GRB spectral evolution 
and pulse shape (e.g. P19), but could also be relevant for the SY emission at lower energies, 
after electron injection stops and the injected electrons migrate toward lower energies, yielding a pulse decay,
provided that $n \leq 1$, because that injected distribution shrinks to quasi mono-chromatic for $n > 1$.

 For the SY spectrum and pulse shape (light-curve) at lower energies (X-ray and optical), we are interested in the 
cooled electron distribution below $\gi$ (or {\sl cooling-tail})
\begin{equation}
 \calN (\gamma < \gi) \sim \gamma^{-m}
\label{Ncool}
\end{equation}
Substitution of that power-law cooling-tail in the conservation Equation (\ref{cons}) leads to $m=n$: 
{\sl the exponent of the cooling-tail} distribution with energy {\sl is equal to the exponent at which the electron energy
appears in the cooling power} for any $n \neq 1$, provided that a certain condition (dependent on the radiative cooling 
process) is satisfied (see P19). Adiabatic cooling, for which $n=1$, does not yield $m=n$. 
A solution-continuity argument (based on the assumption that if the above result $m=n$ is valid for $n > 1$ and $n < 1$, 
then it should also be valid for $n=1$) seems reasonable but is wrong.

\vspace{3mm}
\section{\bf Synchrotron (SY) Cooling}
\label{SY}

 Synchrotron electron cooling is governed by
\begin{equation}
 - \frac{d\gamma}{dt} = \frac{P_{sy}(\gamma)}{m_ec^2} = \frac{1}{6\pi} \frac{\sigma_e}{m_ec} \gamma^2 B^2
\label{Psy}
\end{equation}
where $\sigma_e$ is the cross-section for electron scattering and $B$ is the magnetic field strength.
The photon SY characteristic energy at which an electron radiates is
\begin{equation}
 \esy = \frac{3\,he}{16\,m_e c} B \gamma^2
\label{esy}
\end{equation}
For a constant magnetic field $B$, integration of Equation (\ref{Psy}) leads to the lowest electron energy
\begin{equation}
 \gamma_m(t) = \gi \h \left(1 \h+ \h\frac{t}{\tsyi} \h \right)^{\h-1} , 
 \quad \tsyi \equiv \tsy (\gi) = \frac{\gi m_ec^2}{P_{sy}(\gi)} 
\label{gmsy}
\end{equation}
for an initial electron energy $\gi$, with $\tsyi$ being the SY-cooling timescale for the $\gi$ electrons.
Then, the SY photon energy $\em$ at which $\gm$ electrons radiate and the transit-time $\te$ for a $\gi$ electron 
radiating at the GRB peak-energy $\Ep \simeq 100$ keV to cool to an energy for which its SY characteristic energy 
is $\esy = \eps$ are  
\begin{equation}
 \em (t) = \Ep \left( 1 + \frac{t}{\tsyi} \right)^{-2} \ra \tesy \simeq \epstoEp^{-1/2} \tsyi
\label{tesy}
\end{equation}
For later use, the SY-cooling law of equation (\ref{Psy}) can be written 
\begin{equation}
  - \left( \frac{d\gamma}{dt} \right)_{sy} = \frac{1}{\tsyi}\frac{\gamma^2}{\gi}
\label{sycool}
\end{equation}
and the SY-cooling timescale for an electron of energy $\gamma$ radiating at SY energy $\eps$ is
\begin{equation}
\hh  t_{sy}(\eps) = \frac{\gamma}{-\left(\ds \frac{d\gamma}{dt} \right)_{\h sy}} = \frac{\gi}{\gamma}\tsyi = 
      \epstoEp^{\h -1/2} \h \tsyi = \tesy
\label{tsy}
\end{equation}
Thus, the SY-cooling timescale for an electron of energy $\gamma$ is the transit-time from GRB to the SY 
characteristic energy $\eps(\gamma)$ at which that electron radiates.

\vspace{2mm}
\subsection{\bf Cooled-Electrons Distribution (Cooling-Tail)}

 At $t < \tsyi$, most electrons are at energies above $\gi$ and have a distribution with energy that show the injected one
\begin{equation}
 (t < \tsyi) \quad \quad  \calN (\gi < \gamma) \sim \frac{R_i t}{\gi} \left( \frac{\gamma}{\gi} \right)^{-p} 
\label{pp}
\end{equation}
for a constant injection rate $R_i$. At $\tsyi < t < \ti$, if the magnetic field $B$ is also constant, the cooled 
electron distribution of Equation (\ref{Ntail}) develops, and its normalization at $\gi$ is constant because the number
of electrons above $\gi$ is that injected in the last cooling timescale $\tsyi$, $N (\gamma > \gi) = R_i \tsyi$,
which is constant 
\begin{displaymath}
 (\tsyi < t < \ti \;,\; R_i \sim B^2) :
\end{displaymath}
\begin{equation}
  \calN (\gm < \gamma < \gi) \sim \frac{R_i \tsyi}{\gi} \left( \frac{\gamma}{\gi} \right)^{-2} 
\label{Ntail}
\end{equation}
with the lowest electron energy $\gm m_ec^2$ given in Equation (\ref{gmsy}). 
The above condition for a power-law cooling-tail is satisfied if the magnetic field energy-density ($\sim B^2$) 
is a constant fraction of the internal energy of relativistic electrons ($\sim n'_e \gi$) because the comoving-frame 
density of those electrons should satisfy $n'_e \sim R_i$.

 The growth of the above $\gamma^{-2}$ cooling tail is confirmed by numerically tracking the SY cooling of electrons 
({\bf Figure 1}).

\begin{figure*}
\centerline{\includegraphics[width=16cm,height=12cm]{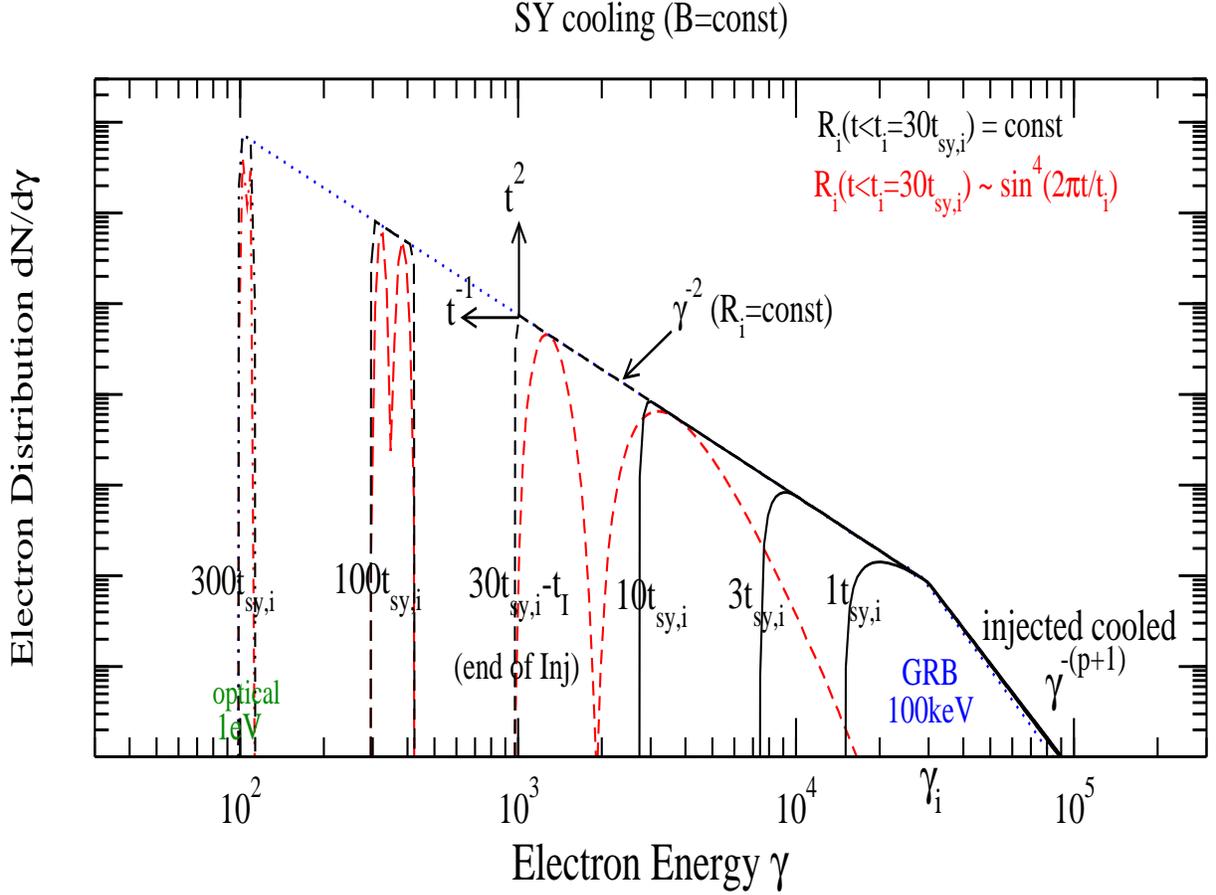}}
\figcaption{\normalsize Evolution of the electron distribution for {\bf synchrotron}-dominated electron cooling
 and for a {\bf constant magnetic field} $B$, obtained by tracking the cooling of infinitesimal electron injections 
 on a (time x energy) grid, using Equation (\ref{sycool}). 
 (This brute-force approach yields more accurate results than a sophisticated integration of the numerically unstable 
 conservation Equation \ref{cons}, and need not be computationally expensive).
 Magnetic field is $B=100$ G, electrons are injected above energy $\gi =3\ttimes 10^4$, with a $p=3$ power-law distribution 
 with energy. For a source Lorentz factor $\Gamma=100$ and redshift $z=1$, the peak-energy of the $\eps \Fe$ spectrum is 
 $\Ep \simeq 100$ keV, the observer-frame SY cooling-time is $\tsyi^{(obs)} = (z+1) \tsyi/(2\Gamma) = 26$ ms.
 for photons emitted by the fluid moving toward the observer (and a factor two larger for those emitted by a region moving 
 at an angle $\Gamma^{-1}$ relative to the direction toward the observer). 
 {\sl The cooled electrons below $\gi$ have a $\gamma^{-2}$ distribution with energy}, as shown analytically (Equation \ref{Ntail}). 
 The lowest electron energy $\gm m_ec^2\sim t^{-1}$ and their corresponding distribution $\calN(\gm) \sim t^2$ satisfy 
 Equations (\ref{gmsy}) and (\ref{Ngmin}).
 After electron injection stops ($t>\ti$), the width of the electron distribution (black lines) narrows as in 
 Equation (\ref{dgamma}) and its {\sl peak slides on the same $\gamma^{-2}$ line} as during electron injection ($t<\ti$). 
 The number of electrons producing SY emission at a given photon energy $\eps$ is constant after the epoch when the $\gamma_m$ 
 electrons "migrate" to that $\eps$ and before the end of energy injection at $\ti$; thus the {\sl SY flux at energy $\eps$ 
 will also be constant}.
 Red lines show a variable electron injection rate $R_i$ that yields two GRB pulses. 
 During their cooling energies, the gap between the two injections decreases and suggests that the GRB variability 
 (with timescale $\ti/2 = 15\,\tsyi$) will be lost in the optical counterpart produced when the GRB electrons reach 
 optically-emitting energies (Panaitescu \& Vestrand 2022). 
}
\end{figure*}

 At $t > \ti >\tsyi$, the electron density at the peak of the cooled-electrons distribution is
\begin{equation}
 \calN (\gm) = \calN (\gi) \frac{d\gi}{d\gm} = \frac{R_i \tsyi}{\gi} \left(\h \frac{\gm}{\gi} \h \right)^{\h -2} \hh
     \simeq \frac{R_i (t+\tsyi)^2}{\gi \tsyi}
\label{Ngmin}
\end{equation}
with $\gamma(t)$ given in Equation (\ref{gmsy}).
Therefore, after electron injection stops, the peak of the cooled-electrons distribution slides on the same cooling curve
$\gamma^{-2}$ ({\bf Figure 1}). 
 The width of the cooled-electrons distribution is
\begin{equation}
  \Delta \gamma (t > \ti) \simeq \frac{R_i \ti}{\calN (\gm)} = \frac{\ti}{\tsyi} \frac{\gm^2}{\gi} \rightarrow 
   \frac{\Delta \gamma}{\gm} = \frac{\ti}{t + \tsyi}
\label{dgamma}
\end{equation}
 Nearly the same result can be obtained easier by using the cooling law of Equation (\ref{gmsy}) to track the evolution 
of the cooling-tail bounds $\gm-\gi$ at $t > \ti$:
\begin{equation}
  \frac{\Delta \gamma}{\gamma_m} (t>\ti) \simeq \frac{\gm(t-\ti)}{\gm(t)} - 1 = \frac{\ti}{t + \tsyi - \ti}
\end{equation}
 Thus, after the end of electron injection, the cooled-electrons distribution shrinks, becoming asymptotically 
mono-energetic at energy $\gm$.

 As shown in {\bf Figure 2} and in figure 2 of P19, if the power-law cooling-tail condition $R_i \sim B^2$ is not 
satisfied, then the cooled-electrons distribution becomes harder if $R_i$ increases or if $B$ decreases faster 
than the power-law condition above. The former case leads to a GRB low-energy slope for the instantaneous spectrum
that is harder than $\bLE = -1/2$ but the latter does not because the decreasing peak-energy $\Ep$ brings at 10 keV 
the high-energy softer SY spectrum.
Conversely, if $R_i$ decreases or if $B$ increases, the distribution of cooled electrons becomes softer, yielding
a spectral slope softer than $\bLE = -1/2$ for the instantaneous spectrum.

\begin{figure*}
\centerline{\includegraphics[width=16cm,height=12cm]{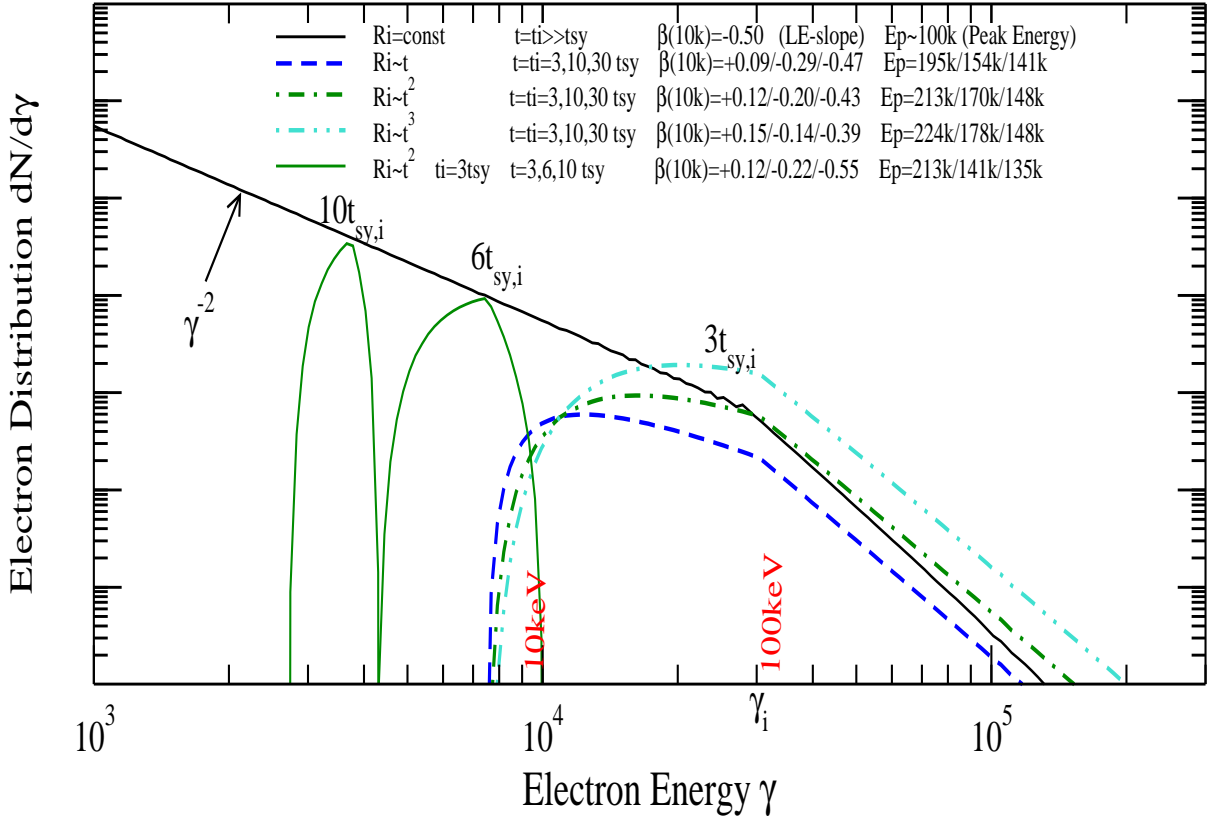}}
\figcaption{\normalsize {\bf Synchrotron}-dominated cooling of electrons injected at an increasing rate $R_i \sim t^y$. 
 Legend indicates the low-energy slope $\bLE$ (at 10 keV) and the peak-energy $\Ep$ (for $B=100$ G, $\Gamma = 100$, $z=1$) 
 of the power-per-decade $\eps \Fe$ at three epochs ($t = 3,10,30\,\tsyi$) for the SY spectrum {\sl integrated} up to those times. 
 {\sl The faster that increase, the harder the cooled-electrons distribution below the typical electron energy $\gi$ and
 the harder is the low-energy SY spectrum}.
 The low-energy spectrum softens progressively, with a hard slope $\bLE \siml 1/3$ at $t < 2\,\tsyi$, 
 an average $\bLE \simeq 0$ at $t \simeq 5\,\tsyi$, and a soft slope $\bLE \siml -1/2$ at $t \simg 30\,\tsyi$, 
 showing how the hardening of $\bLE$ vanishes at $t \gg \tsyi$.
 {\sl Thus, the hardening of the low-energy spectrum for an increasing electron injection rate $R_i$ is a transient feature 
 and spectra with harder slopes $\bLE > 0$ require the cessation of SY emission before the asymptotic $\bLE = -1/2$ is reached}.
 The $\ti = 3\tsyi$ case shows that the softening of low-energy slope $\bLE$ is significantly faster if electron injection stops 
 while the hardening produced by an increasing rate $R_i$ is still effective. }
\end{figure*}

 However, the hardening of the low-energy instantaneous spectrum for an increasing $R_i$ is a {\sl transient} feature 
and disappears after several cooling timescales $\tsyi$ because it depends on the differential/relative time-derivative 
of the injection rate $(dR_i/dt)/R_i \sim y/t$ (for a power-law $R_i \sim t^y$), but lasts longer for faster evolving $R_i$'s, 
as shown by how fast the spectral slopes $\bLE$ given in the legend of {\bf Figure 2} approach the asymptotic value $\bLE = -1/2$. 
For that spectral hardening to become {\sl persistent}, the logarithmic derivative of $R_i$ would have to be constant, 
which means an exponentially-increasing electron injection rate $R_i$.

 Nevertheless, for an increasing rate $R_i$, the hardening of the instantaneous spectrum lasts for a few/several cooling 
timescales $\tsyi$, thus such an $R_i$ yields a GRB low-energy slope for the integrated spectrum harder than $\bLE = -1/2$ 
if the SY emission is integrated over a duration not much longer than $10\,\tsyi$.

\vspace{2mm}
\subsection{\bf Instantaneous Spectrum and Pulse Light-Curve}

\vspace{1mm}
\subsubsection{Pulse Rise}

 The SY spectral peak flux $\fp \sim B N_e \sim B R_i \min(t,\ti)$ at the photon energy $\ep$ where the most numerous 
$\gm$ electrons radiate is
\begin{equation}
  \fp (t) \sim \Fp(\tsyi) \left\{ \begin{array}{ll} t/\tsyi & t < \ti \\ \ti/\tsyi &  \ti < t \end{array} \right. 
\end{equation}
with $\Fp(\tsyi)$ the flux at the peak-energy $\Ep$ of the GRB spectrum.
For a constant on injection rate $R_i$ and constant magnetic field $B$, the flux $\Fp$ increases linearly with time
until $\tsyi$, then remains constant until the end of electron injection at $\ti$ (as indicated by the electron 
distributions of {\bf Figure 1}).
From Equation (\ref{tesy}), the evolution of the spectral peak-energy $\ep \sim B\gm^2$ is approximately
\begin{equation}
  \ep (t) \simeq \left\{ \begin{array}{ll} \Ep & t < \tsyi \\ \Ep (\tsyi/t)^2 & \tsyi < t \end{array} \right. 
\end{equation}
 The SY spectrum at a photon energy $\eps < \Ep$ is
\begin{equation}
  \fe \simeq \fp \left\{ \begin{array}{lll} 
     \hh  (\eps/\ep)^{1/3}  & \eps < \ep   & (t < \te)      \\ 
     \hh  (\eps/\ep)^{-1/2} & \ep  < \eps  & (\te < t < \ti)  \\
           0                & \ep  < \eps  & (\ti + \te < t)  \\
       \end{array} \right.  
\label{sy}
\end{equation}
with $\te$ the epoch when the spectral peak-energy $\ep$ reaches the observing photon energy $\eps$ (Equation \ref{tesy}),
and the last branch due to the exponential cut-off of the SY function. 

 From the above three equations, it follows that, for an electron injection lasting shorter than the SY-cooling timescale 
($\ti < \tsyi$), the pulse light-curve at $\eps < \Ep$ and the {\bf instantaneous} spectrum are
\begin{equation}
 \frac{\fe (t)}{\Fp(\ti)} = \epstoEp^{\h 1/3} \h \left\{ \begin{array}{lll} 
      \ds \frac{t}{\ti}      & \hhhh t < \ti        & \hhh (rise) \\ 
     \hh \left(\ds \h 1 \h + \h \frac{t}{\tsyi}\right)^{\h 2/3}    & \hhhh \ti < t < \tsyi & \hhh (very\; slow\; rise)  \\ 
     \hh  \left( \ds \frac{t}{\tsyi}\right)^{\h 2/3} & \hhhh \tsyi \ll t < \te & \hh (slow \; rise) 
     \end{array} \right.
\label{syLC1}
\end{equation}
where $\Fp(\ti) = \Fp(\tsyi)$ because the GRB peak flux (or flux density at $\Ep$) does not change much from the end
of electron injection at $\ti$ to one SY-cooling timescale $\tsyi$, as there is no significant cooling during that time,
and if the magnetic field is constant.
 This indicates that a low-energy (25-100 keV) GRB pulse should display a very slow rise from the end of electron 
injection at $\ti$ and until the electron SY-cooling timescale $\tsyi$.
 Most GRB pulses are peaky (resembling a double, rising-and-falling exponential or Gaussian -- Norris et al 1996), 
thus the lack of the above slow rise indicates that $\ti \simg \tsyi$, unless the magnetic field evolution shapes 
the pulse rise.

 For an electron injection lasting longer than the SY-cooling timescale ($\ti > \tsyi$) but shorter than the transit-time 
($\ti < \te$) or, equivalently, for a sufficiently low observing energy $\eps < \tilde{\eps} \equiv \Ep (\tsyi/\ti)^2$, 
the pulse light-curve is 
\begin{equation}
 \frac{\fe (t)}{\Fp(\tsyi)} = \epstoEp^{\h 1/3} \h \left\{ \begin{array}{lll} 
     \hh \ds \frac{t}{\tsyi}                    & \hhh t < \tsyi       & \hh (rise) \\ 
     \hh  \left(\ds \frac{t}{\tsyi}\right)^{\h 5/3} & \hhh \tsyi < t < \ti & \hh (fast \; rise)  \\ 
     \hh \ds  \frac{\ti}{\tsyi} \left( \ds \frac{t}{\tsyi}\right)^{\h 2/3} & \hhh \ti < t < \te & \hh (slow \; rise) 
     \end{array} \right.
\label{syLC}
\end{equation}

 Lastly, for an electron injection lasting longer than the transit-time ($\ti > \te$) or for an observing energy 
$\eps > \tilde{\eps}$, the first two rising branches of Equation (\ref{syLC}) remain unchanged (with $\te$ instead 
of $\ti$) and the third rising branch is replaced by a plateau
\begin{equation}
 \fe (\te < t < \ti + \te) = \Fp(\tsyi) \epstoEp^{\h -1/2}  (plateau)
\label{plat}
\end{equation}
for $\te < \ti$.
 The constancy of the SY flux at $t > \te$ is indicated in {\bf Figure 1} by the overlapping cooling-tails. 

 That GRB pulses do not display the plateau expected for $\ti > \te$ indicates that the electron injection timescale 
$\ti$ is not much larger than the transit-time $\te$ from the spectral peak-energy (of the pulse-integrated spectrum) 
$\Ep \simeq 100-200$ keV to an observing energy $\eps = 25-100$ keV : $\ti \siml \te \siml (1-2)\, \tsyi$. 
This conclusion rests on assuming a constant magnetic field and a constant electron injection rate.

 Putting together these two constraints on $\ti$, it follows that the shape of GRB pulses requires that the electron 
injection timescale $\ti$ is comparable to the typical electron SY-cooling timescale $\tsyi$, a conclusion which is
hard to explain. One might speculate that a correlation between $\tsyi$ and $\ti$ could be induced if the injection 
timescale is proportional to the particle acceleration timescale, which for particles accelerated at shocks is 
proportional to the particle gyration timescale; then $\ti \sim \gi/B$. Adding that $\tsyi \sim 1/(\gi B^2)$ points to 
the magnetic field as the reason for a $\ti \sim \tsyi$ correlation; however, the equality $\tsyi \simeq \ti$ would still 
be unexplained.

 Alternatively, the underlying assumption of a constant magnetic field (or varying on a timescale $\tB > \max \{\tsyi,\ti\}$) 
is incorrect. If the magnetic field life-time $\tB < \min \{\ti,\tsyi \}$, then the pulse shape is determined 
by the evolution of $B$, without any relation between the other timescales being implied by GRB observations.

\vspace{1mm}
\subsubsection{Pulse Fall}
\label{pulsefall}

 After the transit-time $\te$ (for $\ti < \te$) or after epoch $\ti + \te$ (for $\te < \ti$), all electrons radiate 
below the observing energy $\eps$, the flux received from the region of angular extent $\Gamma^{-1}$ moving toward 
the observer (the region of maximal relativistic boost $\Gamma$) is exponentially decreasing and the flux received 
becomes dominated by the emission from angles larger than $\Gamma^{-1}$. 
 This "larger-angle emission" (LAE) is progressively less enhanced relativistically and its decay can easily be calculated
if the observer-frame pulse peak-time $\tpk$ is shorter than the angular spread in the photon arrival-time $\tang$. 
In the case of a sufficiently short-lived emission, there is a one-to-one correspondence between the angle of emission 
and the photon arrival-time, so that the LAE decay is (Kumar \& Panaitescu 2000) 
\begin{equation}
  \fe^{(LAE)}(t > \tt_p) = f_{pk} \left( \frac{t}{\tt_p} \right)^{-2+\beta(>\eps)} (fall)
\label{lae1}
\end{equation}
where $\beta (>\eps)$ is the spectral slope at the higher (and higher) photon energy that gets less (and less) Doppler 
boosted to the observing energy $\eps$, 
\begin{equation}
 \hh f_{pk} = \fe (\te) = \left\{ \begin{array}{ll} 
     \hh  \Fp(\ti) & \hhhhh \ti < \tsyi (< \te) \\
     \hh  \ds \Fp (\tsyi) \frac{\ti}{\tsyi} & \hhhhh \tsyi < \ti < \te \\ 
     \hh  \ds \Fp (\tsyi) \epstoEp^{\h -1/2}  & \hhhhh (\tsyi <) \te < \ti 
           \end{array} \right.
\label{fpeak}
\end{equation}
is the pulse peak-flux of Equations (\ref{syLC1}) and (\ref{syLC}), and
\begin{equation}
 \tt_p = \tpk + \tang \;,\; \tang = \frac{R}{2c \Gamma} \;,\; \tpk \simeq \te
\label{tang}
\end{equation}
are the comoving-frame pulse peak epoch, after being stretched linearly\footnote{
 This linearity can be easily proven by calculating the delay in arrival-time between a photon emitted at time $t$
 by the fluid moving directly toward the observer and a photon emitted at time $t + \delta t$ by the fluid moving at angle
 $\Gamma^{-1}$ relative to the direction toward the observer. However, a different recipe for adding timescales will result 
 if times are weighed by the intensity of the emission produced at that time and at a certain angular location.}
by the spread in the photon arrival-time over the region of angular opening $\Gamma^{-1}$, the comoving-frame time-interval 
$\tang$ corresponding to the observer-frame spread in the photon arrival-time $\tang^{(obs)} = R\theta^2/2 = R/2c\Gamma^2$,
and $\tpk$ is the pulse peak-time, as shown by the pulse light-curves given in Equations (\ref{syLC1}) and (\ref{syLC}). 

 If $\tpk \gg \to$ (i.e. for any epoch well after the beginning of electron injection and of the SY emission), 
the integration over the spherical surface up to an angle $\theta = \Gamma^{-1}$ (beyond which the relativistic boost 
${\cal D} = 2\Gamma/(1+\Gamma^2\theta^2)$ decreases substantially) doubles the photon arrival-time $t_0^{(obs)}$ corresponding 
to $\theta=0$ (i.e. from the fluid moving toward the observer). Thus, well after the initial adiabatic timescale, the angular 
integration increases $t_0^{(obs)}$ by 50\% on average and it can be shown that the integration over the spherical surface of 
the photon arrival-time weighed by the received flux yields a relative increase by 1/3.

 For GRB pulses, the peak epoch is $\tpk = \tsyi$ and the peak flux
is $f(\Ep,\tsyi) = \Fp (\tsyi) = \Fp (\ti)$, thus Equation (\ref{fpeak}) {\sl relates} the low-energy pulse peak-flux 
$f_{pk}$ to the flux at the GRB pulse peak $f(\Ep,\tpk)$, which is also the flux at the GRB peak-energy $\Ep$. 
The conclusion that the pulse peak-time $\tpk$ is comparable to the SY-cooling timescale $\tsyi$ is based on the lack of 
slowly-rising and flat-top low-energy GRB pulses expected for a constant magnetic field and a constant electron injection rate. 
If the evolution of these quantities shapes the pulse light-curve, then the pulse-peak epoch is $\tpk = \max \{\ti,\tsyi\}$,
as shown by Equations (\ref{syLC1}) and (\ref{plat}). 

 After noting that the comoving-frame angular timescale $\tang$ is comparable to the AD-cooling timescale 
$\tad = (3/2) t_{co} = (3/2) R/(c\Gamma)$, with $t_{co}$ the comoving-frame ejecta age, the condition that the electron 
cooling is SY-dominated ($\tsyi < \tad$) is equivalent to the angular timescale setting the pulse duration ($\tsyi < \tang$), 
as long as no other factors (duration of electron injection, magnetic field life-time) determine the pulse duration. 
 Thus the pulse rise $t_r$ and fall $t_f$ timescales should always be comparable to $\tang$ and GRB pulses should not be 
too time-asymmetric. Very asymmetric pulses, such as those with a measured ratio $t_r/t_f < 0.2$, require that the emitting 
surface extends much less than $\Gamma^{-1}$, i.e. the pulse emission arises from a bright-spot, and, as shown 
by numerically calculated pulses, a short electron injection timescale $\ti \siml \tsyi/10$ or a magnetic field evolving 
on a timescale $\tB < \tsyi$ are responsible for the asymmetric pulse shape. 

 For GRB pulses, the slope $\beta$ in Equation (\ref{lae1}) is that measured above the peak-energy $\Ep$ but,
for lower-energy (optical and X-ray) pulses, for which the pulse peaks at the transit-time $\te$ when a quasi-energetic
cooled electron distribution "crosses" the observing energy, the above approximation of an infinitesimally short emission 
implies that, after $\tpk = \te$, the pulse turns-off exponentially because there would not be any cooled electrons 
to radiate above the observing energy $\eps$ and whose emission would be (less and less) relativistically boosted to 
energy $\eps$.

 Relaxing the approximation of an infinitesimally short emission, the LAE received after the peak $\tpk = \te$ 
(if $\ti < \te$) or after the plateau-end at $t_{flat} = \te + \ti$ (if $\te < \ti$) will be the integral over the 
ellipsoidal surface of equal arrival-time, with emission from the fluid moving at larger angles relative to the outflow 
origin--observer direction radiating at earlier epochs, when the quasi-monoenergetic cooling-tail was radiating at 
a peak-energy $\em \simg \eps$ (for $\ti < \te$), hence $\beta (\eps) = 1/3$, or when the high-energy end of the 
cooling tail was radiating at $\ep \simg \eps$ (for $\te < \ti$), hence $\beta (\eps) = -1/2$. 

 Then, if the entire surface of the ejecta outflow is radiating at a uniform brightness, the LAE is that given in Equation 
(\ref{lae1}) but with peak-time $\tpk$ stretched by the angular time-spread $\tang$:
\begin{displaymath}
 (fall) \quad  \fe^{(LAE)}(t > \tpk) = 
\end{displaymath}
\begin{equation}
 \hh  \Fp(\tsyi)  \left\{ \begin{array}{ll} \hhh \epstoEp^{\h -5/6} \hh \ds \left(\h \frac{t}{\tsyi} \h \right)^{\h -5/3}
    \h  \left\{ \begin{array}{ll} 1 & \hh \ti < \tsyi (< \te) \\ \hh \ds \frac{\ti}{\tsyi} & \hh \tsyi < \ti < \te 
       \end{array} \right. \\
  \hhh  \ds \epstoEp^{\h -1/2} \h \left(\frac{t}{\ti} \right)^{\h -5/2}  & \hspace*{-27mm} \te < \ti 
    \end{array} \right. 
\label{lae2}
\end{equation}

\vspace{1mm}
\subsubsection{Pulse Light-Curve}

 Equations (\ref{syLC1}), (\ref{syLC}), and (\ref{plat}) provide both the instantaneous spectrum and the {\bf pulse rise} 
or {\bf light-curve} at an energy below gamma-rays (a soft X-ray or optical pulse), for a constant electron injection rate 
$R_i$ and magnetic field $B$, and in the case of a bright-spot emission. The rise is followed by an exponential
decay owing to the electron distribution having cooled to a quasi-monoenergetic one and to the lack of the LAE.
 For a surface of uniform brightness, the same equations give the pulse rise light-curve if timescales are stretched
by the angular time-spread $\tang$, and Equation (\ref{lae2}) gives the pulse power-law decay from the LAE.

 The SY pulse light-curves for SY-dominated electron cooling are also given in Equations (\ref{feic0})--(\ref{feic2})
for iC-dominated electron cooling with an iC-power of exponent $n > 1$ ({\sl Appendix A1}), if one sets $n=2$ and replaces
the iC-cooling timescale $\tic$ with the SY-cooling timescale $\tsyi$.

 The above pulse light-curve equations show that
the optical/X-ray {\sl pulse emission} (instantaneous spectrum) {\sl displays a gradual softening}, with the spectral 
slope 1/3 during the pulse rise evolving to -1/2,-5/6 after the pulse peak. The low-energy slope of GRB pulses softens
from an initial $\bLE = 1/3$ to $\bLE = -1/2$ after $(1-2) \tsyi$, which may explain qualitatively the decrease of 
the count hardness-ratio measured for GRBs pulses (e.g. Bhat et al 1994, Band et al 1997).

\vspace{2mm}
\subsection{\bf Pulse-Duration Dependence on Energy}

 If the pulse duration is set by radiative cooling (Equation \ref{tsy}), then
\begin{equation}
 \dteps = \tsy(\eps) = \epstoEp^{\hh -1/2} \tsyi = \te = \tpk
\label{dtsy}
\end{equation}
with the second to last equality following from (Equation \ref{tesy}) and the last from Equation (\ref{tang}).
The equality of the pulse duration with the pulse-peak epoch stands naturally for any pulse whose rise or fall are not
too fast or too slow, which is the case of the pulse light-curves given in Equations (\ref{syLC1}) and (\ref{syLC}), 
and is an argument which applies to other cooling processes, not just SY. 

 Thus, for SY-dominated electron cooling, the pulse duration should decrease with energy, with the expected 
dependence\footnotemark $\dteps \sim \eps^{-1/2}$ being close to that observed for GRB pulses $\dtg \sim \eps^{-0.4}$. 
\footnotetext{This dependence $\dteps$ derived for energies below the GRB peak-energy $\Ep$, which is assumed constant 
 for the duration of the entire electron injection, applies also for GRB channels above $\Ep$, where the pulse peak epoch
 $\tpk$ is the duration over which electrons accumulate without significant cooling, i.e. $\tpk$ is the SY-cooling 
 timescale for electrons radiating at $\eps > \Ep$.} 
 However, as discussed above, when the electron cooling is SY-dominated ($\tsyi < \tad$), the pulse duration may be set 
by the spread $\tang = \tad/3$ in the photon arrival-time caused by the spherical curvature of the emitting surface
because $\tang > \tsyi$.
 Thus, an immediate consistency between the pulse duration dependence on energy $\dteps$ given in Equation (\ref{dtsy}) 
and GRB observations is readily achieved only if the angular timescale is not dominant, e.g. if the emitting region 
is a small bright-spot of angular extent much less than the "visible" $\Gamma^{-1}$ area moving toward the observer
or if the pulse duration is determined by another timescale (duration of electron injection $\ti$ or magnetic field 
life-time $\tB$) longer than the angular time-spread $\tang$.

 Conversely, for a uniformly-bright spherically-curved emitting surface and for a radiative electron cooling, 
the pulse duration dependence on energy $\dteps$ may be not set by the cooling timescale of that 
radiative process but by the continuous softening of the received emission induced by the differential relativistic boost 
(photons arriving later have less energy) of the emission from the region of angular opening $\Gamma^{-1}$ moving 
toward the observer (corresponding to the pulse rise) and of the larger-angle emission from the fluid outside that
$\Gamma^{-1}$ region (corresponding to the pulse fall).


\vspace{2mm}
\subsection{\bf Pulse-Integrated Synchrotron Spectrum}

 By integrating the above instantaneous spectra over the entire pulse, i.e. past the peak epochs, one obtains the 
pulse-{\bf integrated} spectrum. 
 Due to its fast decay, the contribution of the larger-angle emission is a small fraction of the integral up to the
pulse peak-epoch. 
 For $\te < \ti$, the flat pulse-plateau flux is dominant and trivially sets the slope of the integrated spectrum
$\Fe \sim \eps^{-1/2}$. A more interesting situation occurs for $\ti < \te$, where
\begin{displaymath}
  \Fe (t > \te > \ti) =  \int_0^t \fe(t') dt' \simeq  \int_{\ti}^{\te} \fe(t') dt' 
\end{displaymath}
\begin{displaymath}
  \simeq \Fp(\tsyi) \frac{\ti}{\tsyi} \epstoEp^{1/3} \int_{\ti}^{\te} \left( \frac{t'}{\tsyi} \right)^{2/3} dt' 
\end{displaymath}
\begin{equation}
  \simeq \te \fe(\te) \simeq \ti \Fp(\tsyi) \epstoEp^{-1/2}
\label{intSY}
\end{equation}
with $\fe (\te) = \Fp(\tsyi) (\ti/\tsyi) = \fp$ being equal to the constant flux $\fp (t>\ti)$ at the peak $\ep$ of 
the SY spectrum. (Integrating the instantaneous spectrum only until the pulse peak is a good approximation only for 
the emission from a bright-spot. If the emitting surface is of uniform brightness then, from Equation (\ref{lae2}),
one can show that the post-peak LAE fluence has the same spectrum $\eps^{-1/2}$).

 Thus, although the pulse instantaneous spectrum is hard during the pulse-rise, $\fe (t < \te) \sim \eps^{1/3}$, 
a much softer integrated spectrum $\Fe (t > \te) \simeq [\fe(\te)=const]\, \te \sim \te \sim \eps^{-1/2}$ is obtained
because the transit-time $\te \sim \eps^{-1/2}$ over which the flux is integrated increases with a decreasing energy $\eps$.
Adding that the pulse duration $\dteps$ should be comparable to the transit-time $\te$, the above result suggests that 
the softness of the integrated spectrum can be seen as arising from the pulse duration dependence on the observing energy. 

 Therefore, the pulse-integrated spectrum is $\Fe \sim \eps^{-1/2}$ irrespective of the ordering of electron injection
time $\ti$ and electron transit-time $\te$. This result was derived assuming a constant electron injection rate $R_i$ 
but it is valid even for a variable $R_i$, as shown in {\bf Figure 2} for a power-law electron injection rate. 
 A spectral slope $\bLE =-1/2$ is about half-way on the soft side of the distribution of GRB low-energy slopes. 

 The SY cooling-tail shown in Figure 1 shows the trivial fact that, for a long-lived electron injection, a GRB low-energy 
spectral slope harder than $\bLE = -1/2$ requires that electron cooling or, equivalently, the SY emission stops before 
the GRB-to-10-keV transit-time (Equation \ref{tsy})
\begin{equation}
 \tgxsy = 3\, \E5^{1/2} \tsyi
\label{tgxsy}
\end{equation}
if the electron injection rate $R_i$ is constant, while Figure 2 suggests that the SY emission integrated up to 
$\simeq 2\tgx$ should have a slope $\bLE > 0$ for a rising $R_i(t)$. The same temporal upper limit on the electron cooling 
and SY emission is required by $\bLE > 0$ when electron injection lasts shorter than the GRB-to-10-keV transit-time
as, otherwise, the soft integrated spectrum of Equation (\ref{intSY}) holds. That fact is also illustrated by the spectral 
slopes given in the legend of Figure 2 for the $\ti = 3\tsyi$ case, which shows a soft spectrum if it is integrated
longer than $\ti$.

 Therefore, a harder GRB low-energy slope $\bLE$ requires that the magnetic field fades on a shorter timescale and 
the low-energy slopes of the integrated spectra given in the legend of {\bf Figure 2} suggest that 
\begin{equation}
 ({\bf SY}) \; \bLE = \left\{ \begin{array}{ll} 
      \hh         1/3      & \hh \tB \siml \tsyi \\
      \hh      \in (0,1/3) & \hh \tsyi \siml \tB \siml \tgxsy \\
      \hh     \in (-1/2,0) & \hh \tgxsy \siml \tB \siml 3\, \tgxsy \\
      \hh        -1/2      & \hh 3\, \tgxsy \siml \tB       \end{array}     \right.
\label{bLESY}
\end{equation}


 This {\sl anti-correlation} between the magnetic field life-time $\tB$ and the hardness of the GRB low-energy slope 
applies to any cooling process because it arises from the softening (decrease of peak-energy) of the cooling-tail SY emission.

 The above conclusion that harder GRB low-energy spectral slopes $\bLE$ are the result of electrons {\sl not} cooling 
below the lowest-energy channel (10-25 keV), offers a way to identify GRB pulses arising from bright-spots extending
over much less than the visible region of the ejecta. In absence of a substantial electron cooling and of a significant 
spread in photon-arrival time (due to the small angular extent of a bright-spot), the GRB pulse duration would be more 
time-symmetric at higher energies and their duration should be less dependent on energy.

\vspace{3mm}
\section{\bf Inverse-Compton (IC) Cooling}

 For a constant iC-cooling timescale of the GRB $\gi$-electrons, inverse-Compton ({\bf iC}) cooling is governed by 
\begin{equation}
 - \left( \frac{d\gamma}{dt} \right)_{ic} = \frac{P_{ic}(\gamma)}{m_ec^2} = \frac{1}{\tici} \frac{\gamma^n}{\gi^{n-1}} \;,
 \quad \tici \equiv \tic(\gi)
\label{iCcool}
\end{equation}
with $\tici$ the iC cooling timescale of the GRB $\gi$ electrons.

 If the $\gi$ electrons scatter their own photons in the Klein-Nishina regime ($\gi \Ep' > m_ec^2$), i.e. they cool mostly 
by scattering SY photons $\eps < \Ep$ at the Thomson--Klein-Nishina ({\bf T-KN}) transition, then their cooling begins 
with an index $n = 2/3$ and leads to a cooling tail $\calN (\gamma < \gi) \sim \gamma^{-2/3}$. When the lowest energy 
electrons $\gm$ in the cooling tail begin scattering their own SY photons at the T-KN transition, their cooling exponent 
changes to $n=1$ and a power-law segment of index $m=1$ begins to grow ($\calN (\gm < \gamma) \sim \gamma^{-1}$), 
gradually replacing the pre-existing, higher-energy cooling-tail of index $m=2/3$. When the $\gi$ electrons begin 
to scatter the SY photons produced by the cooling $\gm$ electrons, the entire cooling-tail has index $m=1$ and is again
a single power-law, albeit only until $\tici$ (table 2 of P19).
This $m=1$ cooling-tail arising from iC-cooling dominated scatterings at the T-KN transition has been identified also by 
Nakar, Ando, Sari (2009) and Daigne, Bosnjak, Dubus (2011).

 If the $\gi$ electrons scatter their SY photons in the Thomson regime ($\gi \Ep' < m_ec^2$), their iC-cooling has an index
with $n=\min\{(p+1)/2,2\}$, which changes progressively to $n=\min \{(3p-1)/4,2\}$ and $n=\min\{p,2\}$ (table 1 of P19). 
 
 The iC-cooled electron distribution (i.e. the solution to Equation \ref{cons} for $\calN_i=0$) is a power-law with the
same exponent $-n$ as that of the iC power in Equation (\ref{iCcool}), $\calN (\gamma < \gi) = a(t) \gamma^{-n}$, only 
if $a(t) \simeq \gi^{n-1} R_i \tici = const$, i.e. if $a(t)$ is time-independent. For SY-cooling, this condition 
becomes $R_i \sim B^2$ (Equation \ref{Ntail}), which may have a good reason to be satisfied. For iC-dominated cooling, 
the same condition may be expressed as a relation between $B$, $R_i$ and $\gi$ and has no obvious rationale.

 If the above condition for a power-law cooling-tail is not satisfied, then the cooling-tail should be curved,
with the local slope $n$ depending on the evolutions of the injection rate $R_i$ and magnetic field $B$, which could 
explain why the measured GRB low-energy spectral slopes $\bLE$ have a smooth distribution encompassing the values for
$\bLE = -(n-1)/2$ listed above.

\vspace{2mm}
\subsection{\bf Instantaneous and Integrated Spectra}

 The SY instantaneous spectrum (= pulse light-curve) and integrated spectrum for iC-dominated electron cooling 
are derived in {\bf Appendix A}, where a constant electron injection rate $R_i$ and magnetic field $B$ were assumed.
Then, the condition for the growth of a power-law cooled-electrons distribution, $\tici \sim R_i^{-1}$, 
is equivalent to a constant cooling timescale $\tici$ for the typical GRB electron of energy $\gi$.

 Taken together, these three assumptions can easily be incompatible because the cooling timescale $\tici$
depends on the injection rate $R_i$ and magnetic field $B$ (this is not an issue for SY-dominated cooling 
because, in that case, $\tici$ depends only on $B$). 
Given that the iC-cooling timescale is $\tici \sim \tsyi/Y \sim (B^2 \tau)^{-1}$ with $Y = P_{ic}/P_{sy} \sim \tau$ 
the Compton parameter and $\tau(t) \sim \int_0^t R_i(t') dt'$ the electron optical-thickness to photon 
scattering, a constant $\tici$ requires a decaying magnetic field $B \sim \tau^{-1/2}$ that diverges at $t=0$, 
when the electron injection begins and the optical-thickness is $\tau=0$. 

 It is easy to recalculate the light-curves and spectra that account for an evolving magnetic field $B(t)$, 
which requires to multiply all break energies and spectral peak-flux densities by a factor $B$. However, the evolution
of the magnetic field that ensures the power-law cooling-tail condition $R_i \tici = const$ depends on the 
iC-cooling regime for the $\gi$ electrons (the exponent $n$ of the electron cooling law in Equation \ref{iCcool}), 
thus a generalized treatment is not possible. Furthermore, specializing results to a particular $B(t)$ limits the 
usefulness (if any !) of the results.

 Alternatively, one could assume a constant magnetic field, calculate the time-dependence of the cooling timescale 
$\tici \sim P_{ic}^{-1}$ from the evolution of the iC-cooling power $P_{ic} \sim \tau $, i.e. from the evolution 
of the scattering optical-thickness $\tau$, and integrate the electron cooling law (Equation \ref{iCcool}). However,
the power-law cooling-tail condition $\tici \sim R_i^{-1}$ will not be satisfied (unless a variable $B$ is allowed,
as discussed above for a constant $\tic$) and the SY spectrum above the lowest break-energy will not be a power-law. 
Further use of that essential feature will lead to inaccurate results.

 In conclusion, there is no generalized/comprehensive and accurate way to calculate analytically iC-cooling SY spectra 
and light-curves. We return to all three constancy assumptions (for $R_i$, $B$, $\tic$), and recognize that the 
analytical results of {\sl Appendix A} are only illustrative and of limited applicability.
 
 If the power-law cooling-tail condition is satisfied, then the cooling-tail and its SY emission spectrum are:
\begin{equation}
 \calN (\gamma < \gi) \sim \gamma^{-n} \;,\,  \fe (\eps < \Ep) = \Fp \epstoEp^{-(n-1)/2} 
\end{equation}
the latter result holding for $n > 1/3$ (if $n < 1/3$, the SY emission from the cooled-electrons distribution 
is dominated by the highest energy $\gi$ electrons and is $\fe \sim \eps^{1/3}$, but such a hard cooling-tail is not 
expected to arise). 

 Therefore, the SY {\bf instantaneous} spectrum from the cooling-tail has a low-energy slope $\bLE = -(n-1)/2$. 
The smallest two values for the exponent $n$ of the iC-cooling law, are obtained if the $\gi$-electrons cool weakly 
through scatterings of sub-GRB peak-energy photons at the T-KN transition. For the smallest exponent $n=2/3$, 
the resulting slope $\bLE = 1/6$ is the hardest instantaneous SY spectrum arising from the cooling-tail and the only 
slope harder than the peak of the 
measured distribution $P(\bLE)$. The next exponent $n=1$ allows $\bLE = 0$, which is at the peak of $P(\bLE)$.
All other exponents $n > 1$ occur when the $\gi$-electrons cool strongly by scattering photons in the Thomson regime,
and yield slopes $\bLE < 0$, on the softer half of the measured distribution $P(\bLE)$.

 Equations (\ref{teic1}) and (\ref{teic2}) give the transit-time 
\begin{equation}
 \teic \simeq \tici \left\{ \begin{array}{ll} 
     \epstoEp^{-(n-1)/2} & n > 1 \\ 1 - \epstoEp^{(1-n)/2}  & n < 1 \end{array}  \right.
 \label{teic}
\end{equation}

 For $n > 1$ (electron cooling dominated by iC scatterings in the Thomson regime), integration of the instantaneous 
spectrum over the pulse duration leads to an {\bf integrated} spectrum of similar low-energy slope $\bLE = -(n-1)/2$, 
irrespective of the duration $\ti$ of electron injection relative to the gamma-to-X-ray transit-time $\tgx$, 
therefore GRB low-energy slopes $\bLE > 0$ require that electrons do not cool below 10 keV, i.e. a magnetic field 
life-time $\tB$ shorter than the GRB-to-10-keV transit-time $\tgxic \simeq 3\; \E5^{1/2} \tici$ for $n=2$.
The dependence of the integrated spectrum slope $\bLE$ on the magnetic field lifetime is the same as for SY cooling
(Equation \ref{bLESY}) but with $\tici$ instead of $\tsyi$.

 For $n < 1$ (electron cooling dominated by iC scatterings at the T-KN transition, with only $n=2/3$ possible), 
Equations (\ref{fenless1}) and (\ref{fenless2}) show that the pulse instantaneous spectrum softens progressively 
but the spectral slope of the integrated spectrum is always that of the pulse rise, 1/3 if $\tB < \te$ or 1/6 if $\te < \tB$.
Equation (\ref{intFic}) shows that, even when the soft pulse-decay of spectral slope $-(p-1)/2$ is at maximal brightness 
($\tB > \tpk$, thus the pulse emission is from the $\Gamma^{-1}$ region moving toward the observer), the integrated spectrum 
still has the harder pre-peak slope $1/6$. For a magnetic field life-time $\tB < \tpk$, when the pulse decay is the faster
decaying LAE (because emission from the fluid moving at angles larger than $\Gamma^{-1}$ relative to the observer is less 
beamed relativistically), it is quite likely that the soft pulse-decay contribution to the pulse fluence is dominated by 
the pulse rise.
 Thus, the expected GRB low-energy spectral slope is
\begin{displaymath}
 ({\bf iC/T-KN: \; n < 1}): 
\end{displaymath}
\begin{equation}
 \bLE \h = \h \left\{ \begin{array}{ll} \hh 1/3 & \tB \h < \h \tgxic \\ \hh 1/6 & \tgxic \h < \h \tB \h < \h \ttic \\ 
    \hh 0 & \ttic \h < \h \ti, \tB \end{array} \right.  \tgxic \simeq \frac{1}{3} \tici
\label{bLEIC3}
\end{equation}
with $\ttic$ given in Equation (\ref{ttic}) and with the middle branch second condition ($\tB < \ttic$) being effective
only if $\ttic < \ti$. 

 That the cooling-tail for iC-dominated electron cooling cannot be a perfect power-law, and must have some curvature
(see figure 3 of P19), implies that the actual low-energy GRB spectral slope for iC-cooling through scatterings at 
the T-KN transition spans the range $(0,1/3)$.

\vspace{2mm}
\subsection{\bf Pulse-Duration and Transit-Time}

 Integration of the iC-cooling law of Equation (\ref{iCcool}) allows the calculation of the transit-time to a certain
observing energy and of the pulse duration produced by the passage through the observing band of the SY characteristic 
energy of the electrons that produce the pulse peak. 
For an iC-cooling of exponent $n > 1$ ({\sl Appendix A1}), the pulse peak is set by the passage of the minimal energy of 
the SY spectrum from the cooling-tail, while for $n < 1$ ({\sl Appendix A2}), that epoch is set by the passage of the GRB $\gi$ 
electrons after the end of electron injection at $\ti$, provided that the electron-scattering (optical) thickness is 
approximately constant before $\ti$ (i.e. for a sufficiently fast decreasing electron injection rate) and that the 
magnetic field is also constant.

 For a constant cooling timescale $\tici$, i.e. in the case of a constant magnetic field $B$ and a constant 
electron scattering thickness $\tau$, the pulse duration resulting from the electron iC-cooling is 
\begin{equation}
  \dteps = \frac{\gamma(\eps)}{\ds -\frac{d\gamma}{dt}} = \left(\h \frac{\gi}{\gamma(\eps) \h} \right)^{\h n-1} \hh \tici 
  = \epstoEp^{\h (1-n)/2} \hh \tici 
\label{dtic}
\end{equation}
after using Equation (\ref{iCcool}). 

 For iC-cooling dominated by Thomson scatterings of $\eps \simeq \Ep$ SY photons ($n > 1$), when the rate of electron 
cooling decreases {\sl faster} with decreasing electron energy, {\sl pulses should last longer at lower energy}: 
$\dteps \sim \eps^{-(n-1)/2}$, which is consistent with GRB observations if $n = 2$. Thus, the pulse duration $\dteps$ 
is the same as the transit-time $\te$ (first branch of Equation \ref{teic}). 
 If the electron injection lasts shorter than the transit-time $\ti < \te$, Equation (\ref{fptpic}) shows that the
pulse peak-time $\tpk$ is equal to the transit-time $\te$:
\begin{equation}
 (n > 1) \quad \tpk = \te = \dteps \sim \eps^{(1-n)/2}
\label{tpdtic}
\end{equation}
For $\ti > \te$, the pulse peak is at either $\te$ or $\te + \ti$ depending on the evolution of the injection rate
and of the magnetic field.

 If iC-cooling is dominated by T-KN scatterings ($n < 1$) of $\eps \ll \Ep$ SY photons, then the rate of electron 
cooling decreases {\sl slower} with decreasing electron energy, and {\sl pulses should last shorter at lower energy}: 
$\dteps \sim \eps^{(1-n)/2} = \eps^{1/6}$ (for the one and only $n=2/3$), which is in contradiction with GRB observations
\footnotemark: $\dteps \sim \eps^{-0.4}$. 
\footnotetext{The integration of emission over the equal arrival-time surface may induce a decreasing pulse duration 
  with observing energy, and could reverse the above expected trend, thus this limitation of iC cooling applies mostly 
  to the emission from a bright-spot } 
 The pulse peak-time (Equation \ref{fptpic1}) is set by the transit of the higher energy break $\ep$ after the end 
of electron injection and the pulse duration $\dteps$ is not equal to the transit-time $\te$ (second branch of Equation 
\ref{teic}): 
\begin{equation}
 (n < 1) \quad \tpk = \te + \ti, \; \dteps = \tici - \te
\label{tpdtic1}
\end{equation}

 Further investigations to identify the conditions under which the iC-dominated electron cooling may explain the 
observed trend of GRB pulses to last longer at lower energy are presented in {\sl Appendix A3}.

 The first conclusion is that an increasing scattering optical-thickness $\tau(t)$ affords some 
flexibility to the resulting energy-dependence of the pulse duration $\dteps$ for iC-cooling with $n > 1$ but  
for $n < 1$ pulses should last longer at higher energy, in contradiction with observations.

 The second conclusion is that, for an iC-dominated cooling with $n < 1$, a decreasing magnetic field $B$ should 
lead to a pulse duration dependence on energy that is compatible with observations. Somewhat surprising, the
pulse duration dependence on energy is independent on how fast $B(t)$ decreases, although that result may be an
artifact of some approximations. 
The evolution of $\tau$ does not play any role, however how $B(t)$ and $\tau(t)$ evolve sets the GRB low-energy slope $\bLE$. 

 The above conclusions are relevant for the SY emission from bright-spots of angular opening less than that of the
"visible" region of angular extent $\Gamma^{-1}$, when all pulse properties could be determined by the electron iC-cooling: \\  
$i)$ For electron iC-cooling dominated by scatterings in the Thomson regime ($n \geq 2$), same considerations apply 
 for the pulse time-symmetry and pulse duration dependence on energy as for SY-dominated electron cooling ($n=2$): 
 the faster pulse-rise $t^{2/3}, t, t^{5/3}$ implies a rise timescale $t_r$ that is set by the iC-cooling timescale 
 $\tic$ (Equations \ref{feic0}-\ref{feic2}), while the pulse-fall timescale $t_f$ is set by the pulse peak-time $\tpk$, 
 which is the transit-time $\te \sim \eps^{(1-n)/2}$, thus electron iC-cooling should lead to pulses with a rise-to-fall 
 ratio $t_r/t_f$ that increases with energy, i.e. to pulses which are more time-symmetric at higher energy if pulses 
 rise faster than they fall ($t_r/t_f < 1$), which is in contradiction with observations, but the pulse duration 
 dependence on energy (Equation \ref{dtic}) is consistent with measurements. \\
$ii)$ For iC-cooling dominated by scatterings at the T-KN transition ($n = 2/3$), the pulse-rise $(1-t/\tic)^{-1}, t,
 t/(1-t/\tic)$ is faster than the pulse-fall $(1-t/\tic)^{3(p-1)}$ (Equations \ref{fenless1} and \ref{fenless2}),
 thus a rise-to-fall ratio $t_r/t_f < 1$ independent of energy is expected, which is in accord with observations, 
 but pulses should last longer at higher energy (Equation \ref{dtic}), which is inconsistent with 
 observations. 

 Within the bright-spot emission scenario, the above incompatibilities may be solved by an evolving magnetic field;
alternatively, those incompatibilities disappear if the GRB emission arises from a spherical surface of uniform 
brightness (in the lab-frame), in which case all pulse properties are determined by the spread in photon arrival-time 
and by the emission softening due to the spherical curvature of the emitting surface.

\vspace{3mm}
\section{\bf Adiabatic (AD) Cooling}
\label{AD}

 For a constant radial thickness of the already shocked GRB ejecta, the AD-cooling of relativistic electrons is 
\begin{equation}
 \gm(t) = \gi \left( 1 + \frac{t}{\to} \right)^{-2/3} \hh \ra \; \em (t) \simeq \Ep \left(\frac{t}{\to} \right)^{-4/3}  
\label{gmad}
\end{equation}
with $\em$ the SY characteristic energy $\esy(\gm)$ (assuming a constant magnetic field), thus the AD-cooling 
law is
\begin{equation}
 - \left( \frac{d\gamma}{dt} \right)_{ad} = \frac{P_{ad}(\gamma)}{m_ec^2} = \frac{2}{3} \frac{\gamma}{t+\to} 
\label{adcool}
\end{equation}
and the AD-cooling timescale is
\begin{equation}
  \tad = \frac{\gamma}{\ds -\left( \frac{d\gamma}{dt} \right)_{ad}} = \frac{3}{2} (t+\to)
\label{tad}
\end{equation}
for any electron energy.
 Equation (\ref{gmad}) implies that, at the initial time $\to$ (when electron injection begins), the electron transit-time 
from GRB emission to an observing energy $\eps$ is 
\begin{equation}
  \tead = \to \epstoEp^{-3/4}  
\label{tead}
\end{equation}
for a constant magnetic field.

 Unlike for SY and (most cases of) iC cooling, for AD cooling, where $n = 1$ 
($P_{ad} \sim \gamma$), the conservation Equation (\ref{cons}) does not determine the $\gamma$-exponent of the 
power-law cooling-tail. Instead, that exponent 
can be determined from the continuity of the cooling-tail $\calN (\gamma < \gi) \sim a(t) \gamma^{-m}$ and the cooled 
injected distribution $\calN (\gamma > \gi) \sim A(t) \gamma^{-p}$ at the typical energy $\gi$ of the injected
electrons, where $p$ is the exponent of the injected electron distribution: $\calN_i (\gamma > \gi) \sim R_i \gamma^{-p}$. 

 Substitution of the above two power-law electron distributions in the conservation Equation (\ref{cons}) and the use of 
the AD-cooling law of Equation (\ref{adcool}) lead to
\begin{equation}
 \frac{da}{dt} = - \frac{2}{3}(1-m)\frac{a}{t+\to}  \ra a(t) \sim (t+\to)^{2(1-m)/3}
\end{equation}
\begin{equation}
 \frac{dA}{dt} + \frac{2}{3}(p-1)\frac{A}{t+\to}  \sim R_i(t) 
\end{equation}
\begin{equation}
 R_i \sim (t+\to)^{-y} \ra A(t) \sim (t+\to)^{1-y} 
\end{equation}
where a power-law injection rate $R_i$ was assumed, to allow for an easy solving of the differential equation for $A(t)$. 
The two functions $a(t)$ and $A(t)$ are continuous at $\gi$ only if they have the same time-dependence, which implies that
\begin{equation}
 m = \frac{1}{2} (3y-1) 
\label{mm}
\end{equation}
thus, the slope of the cooling-tail depends on the evolution of $R_i$. The slope of the cooling-tail {\bf instantaneous}
SY spectrum, $\beta = d \ln \fe /d \ln \eps = \min[1/3,-(m-1)/2]$, is 
\begin{equation}
 \bLE = \left\{ \begin{array}{lll} 1/3 & (y < 5/9, &  m < 1/3) \\ 3(1-y)/4 & (y > 5/9, & m > 1/3) \end{array} \right.
\label{betaAD}
\end{equation}
For $y > 5/9$, the cooling-tail SY spectrum becomes softer for a faster-decreasing injection rate $R_i$;
for $y < 5/9$, the cooling-tail is harder than $\calN (\gamma < \gi) \sim \gamma^{-1/3}$ and its SY emission is overshined 
by that from the highest-energy $\gi$ electrons in the cooling-tail, leading to a hard $\bLE = 1/3$ spectrum. That is the 
case for a constant $R_i$: $y=0 \rightarrow m=-1/2$.

 Equations (\ref{FAD1}) and (\ref{FAD2}) of {\bf Appendix B} show that the instantaneous spectrum of AD-cooling 
electrons is harder during the pulse rise than during the pulse fall, with the pulse peak occurring at the time 
$\te$ (if $y > 1$) when the photon energy $\em \equiv \esy(\gm)$ crosses the observing energy $\eps$ or at the time 
$\tp$ (if $y < 1$ - Equation \ref{tpad}) when the higher break-energy $\ep$ of the last injected $\gi$-electrons 
crosses $\eps$. For GRB spectra at the lowest observing energy (10 keV), these crossing epochs are
\begin{equation}
 \tgxad \simeq 5.6 \E5^{3/4} \to \;, \quad \tpxad = \frac{\ti}{\to} \tgxad > \tgxad
\label{tgxad}
\end{equation}
 Equation (\ref{FintAD}) shows that the SY spectrum {\bf integrated} over the entire pulse has a soft slope $\bLE = -3/4$ 
(if the injected electron distribution has an index $p > 5/2$), being softer than that of the instantaneous spectrum 
(Equation \ref{betaAD}) for a reason similar to that discussed above for the integrated spectrum from SY-cooling electrons. 

 Consequently, for the integrated spectrum of AD-cooling electrons to display a hard low-energy slope, 
the instantaneous spectrum must not be integrated past the crossing epochs $\tgx$ and $\tpx$, i.e. the SY emission
must stop and the magnetic field must disappear before the pulse-peak epochs $\tpk$ given in Equation (\ref{fptpad}): 
\begin{equation}
 (R_i \sim t^{-y}): \; \bLE = \frac{1}{3} \rightarrow \tB < \left\{ \begin{array}{lll} \hh \tgxad & \hh (y > 5/9) \\ 
   \hh  \tpxad & \hh (y < 5/9) \\ \end{array} \right.
\label{bLEAD13}
\end{equation}
 The epoch $\tB$ when the magnetic field fades out is before the natural pulse-peak, thus $\tB$ becomes the pulse 
peak-epoch, after which the LAE emission describes the pulse decay, and the pulse duration $\dteps$ has a weaker 
dependence on $\eps$ than given below.

 If the magnetic field lives longer than the crossing time $\tgxad$, then a softer spectrum results after the crossing 
of the lower-end energy $\em$ of the cooling-tail SY spectrum
\begin{displaymath}
 (5/9 < y < 2) 
\end{displaymath}
\begin{equation}
 \hh  \bLE = \frac{3}{4}(1-y) \in \left(\h -\frac{3}{4}, \frac{1}{3} \right) \rightarrow \tgxad < \tB < \tpxad 
\label{bLEAD}
\end{equation}
and an even softer integrated spectrum is produced by the passage of the higher-end energy $\ep$ of the cooling-tail 
SY spectrum
\begin{equation}
  \bLE = -3/4 \ra \tB > \left\{ \begin{array}{lll} \tgxad & (y > 2) \\ \tpxad & (p > 5/2) \\ \end{array} \right.
\label{bLEAD34}
\end{equation}
with $-p$ the exponent of the injected electron distribution with energy. 
 For an injected distribution with $p < 5/2$, the integrated spectrum is dominated by the emission from the 
injected electrons, as they cool after the end of electron injection, with AD-cooling preserving the slope of their 
distribution with energy: $\calN (\gp < \gamma) \sim \gamma^{-p}$, thus $\bLE = -(p-1)/2 \in (-1/2,-3/4)$ for 
$p \in (2,2.5)$ is harder than for the last case above.
 
 If the magnetic field lasts longer than the pulse-peak epoch $\tpk$ given in Equation (\ref{fptpad}), then the pulse 
duration corresponding to the cooling-law given in Equation (\ref{adcool}) is 
\begin{equation}
  \dteps =  \frac{\gamma(\eps)}{\ds - \frac{d\gamma}{dt} }(t=\tpk) \simeq \frac{3}{2} \tpk \simeq 
   \epstoEp^{-3/4} \left\{ \begin{array}{ll} \hh \ti & y < 1 \\ \hh \to & y > 1  \end{array} \right.
\label{dtad} 
\end{equation}
 Thus, AD-dominated electron-cooling should yield pulses whose duration decreases with the observing energy $\eps$, 
as is observed, but the resulting dependence $\dteps \sim \eps^{-3/4}$ is stronger than measured.
However, figure 5 of P19 shows that the numerically-calculated pulses display a duration dependence on energy that 
is weaker than in Equation (\ref{dtad}) and consistent to that measured.

 That the comoving-frame angular time-spread $\tang = R/(2c\Gamma)$ over the visible $\Gamma^{-1}$ region of maximal 
relativistic boost (by a factor $\Gamma$) is always 3 times smaller than the current comoving-frame adiabatic 
timescale $\tad = 1.5\,t = 1.5\,R/(c\Gamma)$, implies that, for AD-dominated electron cooling, all pulse 
properties are determined by the electron cooling and the above pulse duration dependence on energy is accurate for 
either a bright-spot emission or a uniform brightness surface.

\vspace{3mm}
\section{\bf Synchrotron and Adiabatic Cooling}
\label{ADSY}

 Equations (\ref{sycool}) and (\ref{adcool}) show that the SY and AD cooling powers are equal at the {\sl critical} 
electron energy
\begin{equation}
 \hh \gc = \frac{2\tsyi}{3(t+\to)} \gi \rightarrow \left\{ \begin{array}{lll} 
  \hh \gamma \h <\h  \gc, & \hh P_{ad} \h >\h  P_{sy} & \hh {\rm (AD\h-cool)} \\ 
  \hh \gc \h <\h  \gamma, & \hh P_{ad} \h <\h  P_{sy} & \hh {\rm (SY\h-cool)} 
   \end{array}  \right.
\label{gcr}
\end{equation}
 Below the critical electron energy $\gc$, electrons cool adiabatically and the slope of the cooling-tail 
$\calN (\gm < \gamma < \gc)$ is determined only by the history of the electron injection rate $R_i$. Above $\gc$, 
electrons cool radiatively and the slope of the cooling-tail $\calN (\gc < \gamma < \gi)$ is set by the history of 
the electron injection rate $R_i$ and of the magnetic field $B$ (which sets the radiative cooling power).

 At $t=0$, the typical $\gi$ electrons cool adiabatically if $3\to < 2\tsyi$ and radiatively if $2\tsyi < 3\to$.
{\bf Appendix C} shows that the solution (Equation \ref{syadsol}) to the AD+SY electron cooling implies that, 
if the $\gi$ electrons are initially cooling adiabatically ($\gi < \gc(t=0))$, then their cooling remains adiabatic 
all times ($\gm(t) < \gc(t)$, with $\gm(t=0) = \gi$), while if the $\gi$ electrons are initially cooling radiatively 
($\gc(t=0) < \gi$), then their cooling switches from radiative to adiabatic after a "critical" time $t_c$ 
(Equation \ref{tgcr}) defined by $\gc(t_c) = \gm(t_c)$. Thus, in either case, the electrons {\sl cool adiabatically 
eventually}, yet the exact electron cooling law (Equations \ref{exp1}-\ref{exp3}) is close to (1/3 of) that expected 
for SY-dominated cooling: $\gm(t) \sim t^{-1}$ (Equation \ref{gmsy}). 
 
 It may be surprising that, if SY and AD electron-cooling are considered separately, they lead to the opposite conclusion.
The timescales for these two cooling processes, given in Equations (\ref{tsy}) and (\ref{tad}), indicate that both 
cooling timescales increase linearly with time, but faster for AD-cooling ($t_{ad} \simeq 1.5 t$) than for SY 
($t_{sy}(\gm) \simeq t$). Consequently, if the $\gi$ electrons begin by cooling radiatively ($\tsyi < \tad (t=0) = 
1.5\,\to$), then $\tsy (\gm) < \tad$ at any time, thus the electrons cool radiatively at all times.
 Conversely, if the $\gi$ electrons cool adiabatically initially ($1.5\,\to < \tsyi$), then their cooling switches 
to SY-dominated at a (erroneous) critical time $t_{cr}$ defined by $\tsyi [\gm(t_{cr})] = \tad (t_{cr})$ (which leads to 
$t_{cr} = 2\tsyi - 3\,\to$), after which $\tsy [\gm(t)] < \tad$ and the electron cooling should become SY-dominated.

 Thus, if the two electron cooling processes are treated as acting independently, the electron cooling becomes radiative 
at late times irrespective of which cooling process was dominant initially, in total contradiction with the expectations
from the solution to the double-process cooling, which shows that electron cooling should always become adiabatic eventually. 
 The reason for this discrepancy is the unwarranted (ab)use of the SY-cooling solution (Equation \ref{gm}) in the calculation 
of the SY-cooling timescale (Equation \ref{tsy}), which is correct only at early times and only if the electron-cooling 
begins in the SY-dominated regime, but is incorrect at later times, when the SY and AD cooling timescales $\tsy[\gamma(t)]$ 
and $\tad \simeq t$ are comparable and when the exact electron-cooling law (Equation \ref{syadsol}) is inaccurately described 
by the SY-cooling of Equation (\ref{gmsy}).

 Despite this fundamental differences in the expectations for the single- and double-process cooling, the asymptotic 
SY solution at late times over-estimates the exact electron energy only by a factor up to 3. Thus, if one makes the mistake 
of using the SY-cooling solution whenever that process appears dominant, the resulting error is an over-estimation by up to 
an order of magnitude of the corresponding spectral break energies and by up to a factor 3 of the corresponding transit-times.
 
 The upper limits on the magnetic field life-time $\tB$ given in Equation (\ref{bLEAD13}) are valid if the cooling of 
the lowest-energy $\gm$ electrons (for $y > 5/9$), or of the GRB $\gi$ electrons after the end of electron injection 
(for $y < 5/9$), is described by the AD-cooling solution of Equation (\ref{gmad}) until the corresponding transit-times 
given in Equation (\ref{tgxad}). 
 If the electron cooling is AD-dominated initially ($\to < \tsyi$), then it remains so at any later time. However, 
the AD-cooling law of Equation (\ref{gmad}) remains valid only until the switch-time $\tt$ defined in Equation (\ref{tt}), 
after which the electron cooling is described by the 1/3-SY solution, even though the electron cooling is AD-dominated. 
 Thus, the results for the GRB low-energy slope $\bLE$ of \S\ref{AD} are applicable if the crossing-times $\tgxad$ and 
$\tpxad$ are shorter than the switch-time $\tt$, which lead to the same restriction: $\to,\ti < 0.2\,\tsyi$. 

 Therefore, AD-cooling sets alone the GRB pulse light-curve and integrated spectrum if the radiative (SY) cooling timescale 
is at least an order of magnitude larger than the initial ejecta age $\to$ and the duration $\ti$ of electron injection.
 The evolution of the electron distribution undergoing AD and SY cooling, with the strength of AD cooling increasing from 
$\to = 0.1 \tsyi$ to $\to = 0.01 \tsyi$, is shown in {\bf Figure 3} and supports the above conclusion.

\begin{figure*}
\centerline{\includegraphics[width=16cm,height=13cm]{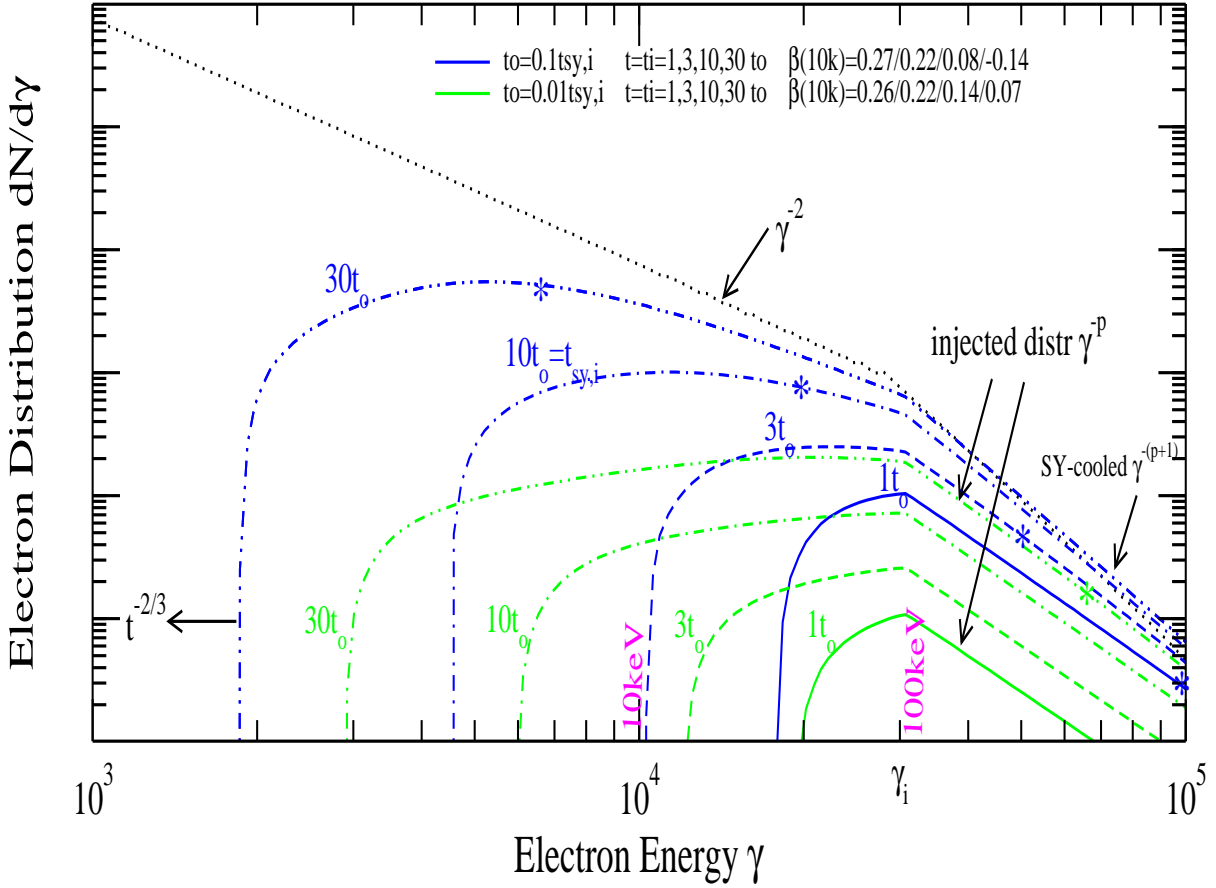}}
\figcaption{\normalsize {\bf Adiabatic} and {\bf Synchrotron} cooling, for the same parameters as in Figure 1
 but for a stronger AD cooling of increasing strength: initial timescale (initial ejecta age) $\to= 0.1\tsyi$ (blue) 
 and $\to = 0.01\tsyi$ (green). {\sl For a shorter initial time $\to$, the AD-cooling tail is more extended}.
 Electron injection lasts until after the spectrum integration epochs ($\ti=t$). 
 Stars indicate the critical electron energy $\gc$ for which AD and SY losses are equal. 
  For $\to \siml 0.1 \tsyi$, the lowest energy $\gm$ electrons cool adiabatically at all times and their energy decreases 
 as in Equation (\ref{gmad}) at $t < \tt$ (Equation \ref{tt}); at $t > \tt$, their cooling turns asymptotically to the 
 (1/3)-SY cooling solution discussed in {\sl Appendix C}, which is just that: 1/3 of the energy given in Equation (\ref{gmsy}) 
 for synchrotron cooling.
  Electrons with energy $\gamma < \gc$ cool adiabatically, leading to a cooled-electrons distribution $\calN (\gamma < \gc) 
 \sim \gamma^{1/2}$ (for a constant electron injection rate $R_i$) and to a hard $\fe \sim \eps^{1/3}$ SY spectrum, 
 while those with $\gamma > \gc$ cool radiatively, leading to $\calN (\gc < \gamma < \gi) \sim \gamma^{-2}$ 
 (for a constant $R_i$ and magnetic field $B$).
 At $t < 2\,\tsyi$, the $\gx$ electrons radiating at 10 keV (mid X-rays) are below the critical energy ($\gc > \gx \simeq \gi/few$) 
 and their adiabatic-dominated cooling leads to a persistent GRB low-energy spectral slope $\bLE > 0$, with softer slopes 
 $\bLE$ resulting for a decreasing $R_i(t)$ (Equation \ref{bLEAD}). 
 At $t > 2\,\tsyi$, when $\gc < \gx$, the synchrotron-dominated cooling of the mid-X-ray radiating electrons leads to 
 {\sl a progressive softening of the instantaneous spectrum at 10 keV, asymptotically reaching the expected slope $\bLE = -1/2$}.
 Some of that softening is captured in the integrated spectrum slopes $\bLE$ listed in the legend. 
 Indicated photon energies are for $z=1$.
}
\end{figure*}

 The low-energy slope $\beta$ of the GRB instantaneous spectrum depends on the location of the SY characteristic energy 
$\esy(\gc)$ relative to the lowest-energy channel (10 keV) of GRB measurements or, equivalently, the location of the electron 
critical energy $\gc$ relative to the energy $\gx \sim \gi/3\E5^{1/2}$ of the electrons that radiate at 10 keV. 
 From Equation (\ref{gcr}), it follows that, if the $\gi$-electron cooling begins AD-dominated ($\to \ll \tsyi$),
then their cooling remains AD-dominated (i.e. $\gamma_c > \gi$) until epoch $t = (2/3)\tsyi$ and the cooling 
of $\gx$-electrons remains AD-dominated until epoch $t \simeq 2\,\E5^{1/2} \tsyi$.
 For $t > 2\,\tsyi$, SY-cooling sets the cooling-tail $\calN (\gx < \gamma < \gi)$ energy distribution below the 
GRB peak-energy $\Ep$, leading to a softening of the SY emission to the expected asymptotic slope $\bLE = -1/2$.
 The legend of {\bf Figure 3} shows that {\sl expected gradual softening of the instantaneous GRB low-energy slope}. 

 Thus, the condition for AD-cooling to set the low-energy GRB spectral slope leads to an upper limit on the 
magnetic field life-time: $\tB < (2/3-2)\tsyi$, to switch-off the soft SY emission at 10-100 keV produced
by the soft cooling-tail above $\gc$ resulting through SY-dominated electron cooling.
This condition on $\tB$ is satisfied on virtue of Equations \ref{bLEAD13} and \ref{bLEAD13}) if electron cooling 
begins well in the AD-dominated regime $\to \ll \tsyi$ and if $\ti \ll \tsyi$.

\vspace{3mm}
\section{\bf Discussion}

\vspace{1mm}
\subsection{\bf GRB Low-Energy Slope $\bLE$ for SY Cooling}

 Equation (\ref{bLESY}) and numerical calculations (Figure 2) show that a low-energy (10 keV) GRB spectral slope 
$\bLE < 1/3$ of the pulse-{\sl integrated} spectrum results from an {\sl incomplete/partial electron cooling} due
to the magnetic field life-time $\tB$ being comparable to the GRB-to-10-keV transit-time $\tgxsy$ (Equation \ref{tgxsy}) 
that it takes the typical GRB electron (radiating initially at the GRB spectrum peak-energy $\Ep \sim 100$ keV) 
to cool to an energy for which the corresponding SY characteristic photon energy is 10 keV. More exactly, a slope
$\bLE \siml 1/3$ results for $\tB < 2\,\tsyi$, $\bLE \simeq 0$ requires that $\tB \simeq (3-5)\,\tsyi$, and $\bLE \simg -1/2$ 
is obtained for $10\,\tsyi \siml \tB$. 

 For a constant electron injection-rate $R_i$ and magnetic field $B$, SY cooling over a duration longer than 
$3\,\tsyi$ leads to a soft slope $\bLE = -1/2$, irrespective of the duration $\ti$ over which electrons are injected: \\
$i)$ For $\tsyi < \ti,\tB$, the electron distribution develops a cooling-tail with energy distribution 
  $\calN (\gamma < \gi) \sim \gamma^{-2}$ at $t > \tsyi$, for which the SY instantaneous spectrum is 
  $\fe \sim \eps^{-1/2}$ and the integrated spectrum is the same. That GRB pulses do not have a flat plateau at
  their peak, starting at the transit-time $\tgxsy$ and until the end of electron injection at $\ti$, indicates
  that either $R_i$ or $B$ are not constant,  \\
$ii)$ For $\ti < \tsyi < \tB$, a power-law cooled electron distribution does not develop; instead that distribution 
  shrinks to a mono-energetic one after $\tsyi$. Integration of the SY instantaneous spectrum $\fe (\eps < \ep) 
  \sim \eps^{1/3}$ until after the SY characteristic energy $\ep$ at which the cooled GRB electrons radiate decreases 
  below 10 keV leads to an integrated spectrum with the same low-energy slope $\bLE = -1/2$ as for a cooling-tail. 

 This coincidence arises from that a cooling-law $d\gamma/dt \sim \gamma^{-n}$ yields 
$i)$ a cooling-tail $\calN \sim \gamma^{-n}$ whose SY spectral slope is $\beta = -(n-1)/2$ and 
$ii)$ an electron cooling $\gamma \sim t^{-1/(n-1)}$ for $n > 1$, a transit-time $\te \sim \eps^{-(n-1)/2}$ and
   an integrated spectrum $\Fe \simeq \fe(\te)\te \sim \te \sim \eps^{-(n-1)/2}$.

 SY electron cooling can yield cooling-tails harder (softer) than $\calN (\gamma < \gi) \sim \gamma^{-2}$ and corresponding 
SY spectra harder (softer) than $\bLE = -1/2$ if electrons are injected at an increasing (decreasing) rate $R_i$. 
Because the hardness of the 10-100 keV SY spectrum is set by the electrons injected during the last few cooling timescales 
($\tgxsy \sim 3\,\tsyi$), a variable electron injection rate $R_i$ can change the resulting cooling-tail only if the 
injection rate variability timescale is shorter than the transit-time $\tgxsy$. This means that an electron injection rate 
$R_i$ that varies as a power-law in time, and which has a variability timescale equal to the current time, can alter the 
cooling-tail index only over a duration comparable to transit-time $\tgxsy$. 
 Conversely, an electron injection rate $R_i$ that is a power-law in time does not change significantly over the 
second $\tgxsy$ and leads to the standard slope $\bLE = -1/2$. 
 Consequently, a variable electron injection rate $R_i$ can change the above magnetic field life-times $\tB$ by a factor up 
to two. 

 Harder (softer) cooling-tails can also be obtained if the magnetic field $B$ decreases (increases), but a change in the 
low-energy slope $\bLE$ of the pulse-integrated spectrum is less feasible because a decreasing $B$ leads to a decreasing
SY spectrum peak-energy $\Ep$ which compensates the effect that a decreasing magnetic field has on the hardness of the 
cooling-tail, while an increasing $B$ could lead to an increasing peak-energy $\Ep$ that is in contraction with observations.

 Thus, there is a direct {\sl mapping} between the distribution of the GRB low-energy slope $P(\bLE)$ and that of the 
magnetic field life-time $P(\tB)$. The peak of the slope distribution $P(\bLE)$ at $\bLE = 0$ implies that the life-time 
distribution $P(\tB)$ peaks at $\tB \simeq 3\,\tsyi$, which means that the {\sl generation of magnetic fields in GRB 
ejecta is tied to the cooling of the relativistic electrons}. 

 The puzzling feature of the $\bLE - \tB$ correlation is that the distribution of slopes $\bLE$ does not exhibit peaks 
at the extreme values $\bLE = 1/3$ (corresponding to $\tB < \tsyi$) and $\bLE = -1/2$ (corresponding to 
$\tB > 10\,\tsyi$), which may be explained in part by the statistical uncertainty $\sigma (\bLE) \simeq 0.1$ of measuring 
the GRB low-energy slope $\bLE$.

\vspace{2mm}
\subsection{\bf GRB Low-Energy Slope $\bLE$ for AD Cooling}

 For AD cooling, the cooling-tail distribution is determined by the only factor at play, the electron injection rate, 
assumed here to be a power-law in time $R_i \sim t^{-y}$, which provides all the flexibility needed, but using only one parameter.

 In contrast to SY-dominated electron cooling, where the dependence on the injection rate $R_i$ of the cooling-tail distribution 
is a{\sl transient} feature, lasting for a few SY-cooling timescales $\tsyi$, the power-law cooling-tail resulting for 
AD-cooling is a {\sl persistent} feature because a substantial change in the rate $R_i$ is guaranteed to occur during an AD-cooling 
timescale, given that both timescales are the same (the current time).
 Similar to SY-cooling, for AD-dominated electron cooling, the passage of the peak-energy of the SY spectrum from the
cooling-tail leads to a softer integrated spectrum with $\bLE =-3/4$.

 Consequently, AD-cooling allows easier than SY-cooling a range of spectral slopes for the instantaneous spectrum.
That diversity is imprinted on the integrated spectrum if the cooling-tail contribution is dominant (which requires $y < 2$) 
and if the magnetic field has a life-time $\tB$ between the transit-times $\te$ (Equation \ref{tead}) and $\tp$ 
(Equation \ref{tpad}) corresponding to the low and high-energy ends $\em$ and $\ep$ of the cooling-tail spectrum 
crossing the observing energy. 

 Equation (\ref{bLEAD}) for the GRB low-energy slope $\bLE(y)$ shows that the measured distribution of the GRB slope 
$\bLE \in (-3/4,1/3)$ (which most of the range of GRB slopes) maps directly to the distribution of the exponent 
$y \in (5/9,2)$ of the electron injection rate $R_i \sim t^{-y}$ -- $P(y) = (3/4) P(\bLE)$ -- for a magnetic field life-time 
$\tB \in (\tgxad,\tpxad)$ (Equation \ref{tgxad}). 
 For a $\tB$ outside the above range, the GRB low-energy slope can be a hard $\bLE = 1/3$ (Equation \ref{bLEAD13}) or a soft 
$\bLE = -3/4$ (Equation \ref{bLEAD34}), with even softer slopes $\bLE < -3/4$ occurring if the integrated spectrum is dominated 
by the SY emission from GRB electrons of energy above $\gi$. 

 As for SY-dominated electron cooling, this conclusion comes with two puzzles: it implies a correlation between 
the magnetic field life-time $\tB$ and the cooling of the lowest and highest-energy electrons in the cooling-tail 
via the GRB-to-10-keV transit-times (Equation \ref{tgxad}), and peaks in the $P(\bLE)$ distribution at $\bLE = 1/3$ and $\bLE = -3/2$.

\vspace{2mm}
\subsection{\bf GRB Low-Energy Slope $\bLE$ for iC Cooling}

 If the typical GRB electrons of energy $\gi$ cool through scatterings (of SY photons produced same electrons) 
in the {\sl Thomson} regime ($\gi \Ep' < m_e c^2$), when the cooling-power exponent is $n \geq 2$, the integrated spectrum 
shows the same features and dependence on the magnetic field life-time $\tB$ as for SY-dominated electron cooling 
(for which $n=2$): \\
$i)$ crossing of the lowest-energy of the cooling-tail SY spectrum softens the integrated spectrum to the slope 
  $\bLE = -(n-1)/2$, whether or not the electron injection lasts longer than the GRB-to-10-keV transit-time 
  $\tgxic \simeq 3\,\tici$, i.e. whether the cooling-tail develops down to an energy for which the SY characteristic 
  energy is below 10 keV or shrinks to a monoenergetic distribution before reaching the observing energy, \\
$ii)$ hard GRB low-energy spectra require an incomplete electron cooling due to a short-lived magnetic field, lasting 
  about the transit-time $\tgxic$ (Equation \ref{teic1}), and there should be a one-to-one correspondence between the
  GRB low-energy slope $\bLE$ and the magnetic field life-time $\tB$, modulo a possible variation of the electron
  injection rate $R_i$, whose effect lasts only for about $\tgxic$, with $\tB \simeq \tgxic$ accounting for the peak of 
  the measured $P(\bLE)$ distribution at $\bLE = 0$.

 The iC-cooling of GRB electrons through scatterings at the {\sl T-KN transition} ($\gi \Ep' > m_e c^2$), when $n=2/3$, has \\
$i)$ a similarity with the AD-dominated electron cooling ($n=1$) in that an energy-wide cooling-tail persists after the 
end of electron injection, \\
$ii)$ a similarity with the SY-dominated electron cooling ($n=2$) in that the crossing of either end of the cooling-tail 
(at $\tgxic$ or at the pulse-peak epoch $\tpk =\tgxic+\ti$) yields an integrated spectrum with the same slope $\bLE = 1/6$ 
as for the SY emission from the cooling-tail, and \\
$iii)$ a unique feature in that, after the time $\ttic$ of Equation (\ref{ttic}), the cooling-tail of exponent $n=2/3$
is replaced by one with $n=1$, provided that the electron injection lasts $\ti > \ttic$, which leads to an instantaneous 
SY spectrum of slope $\beta = 0$ that yields an integrated spectrum of slope $\bLE = 0$ (which is the peak of the measured
low-energy slope distribution - Equation \ref{PbLE}), if the magnetic field life-time satisfies $\tB > \ttic$.  

 IC-dominated electron cooling with $n=2/3,1$ cannot lead to integrated spectra with a low-energy slope $\beta < 0$
because the contribution to the integrated spectrum from the GRB electrons above $\gi$ is smaller than that from the 
cooling-tail after the end of electron injection. 
 Thus, one important feature of electron-cooling dominated by iC-scatterings at the T-KN transition (with $n \leq 1$) 
is that, without the diversity in slopes $\bLE$ allowed by a variable electron injection rate, it can explain {\sl only} 
the harder half of the measured distribution of GRB low-energy slopes, with $\bLE \geq 0$ 
(Equation \ref{bLEIC3}): the hardest slope $\bLE = 1/3$ requires that $\tB < \tgxic$ and the slope $\bLE = 0$ at the 
peak of the $P(\bLE)$ distribution requires that $(\tB,\ti) > \ttic$. 

 {\it However, that electron cooling dominated by iC-scatterings in the Thomson regime yields a GRB low-energy slope $\bLE = -1/2$
while iC-cooling dominated by scatterings occurring at the T-KN transition yields a persistent slope $\bLE = 0$ suggest 
that diversity among bursts in the scattering regime that dominates the iC-cooling may yield intermediate slopes $\bLE$.
To that end, Daigne, Bosnjak, Dubus (2011) have illustrated how the transition from a soft low-energy spectrum to a 
harder one is obtained by 
$i)$ replacing SY-cooling ($Y(\gi) < 1$) or iC-cooling in the Thomson regime ($\gi \Ep' < mc^2$, $Y(\gi) > 1$) with 
  iC-cooling at the T-KN transition ($\gi \Ep' > mc^2$, $Y(\gi) > 1$) and by 
$ii)$ increasing the Compton parameter, leading to $\bLE (Y<1) = -1/2$ to $\bLE (Y\gg 1)=0$. }

\vspace{3mm}
\section{\bf Conclusions}

 The aim of this work is to examine the implications of the low-energy slopes $\bLE$ measured for GRBs by CGRO/BATSE 
and Fermi/GBM within a simple model where relativistic electrons (of typical energy $\gi m_e c^2$) in a magnetic 
field ($B$) produce SY emission in a relativistic source (of Lorentz factor $\Gamma$) and at some radius ($R$).

{\sl Low-energy slope of instantaneous SY spectrum}.
 That slope depends on the dominant electron-cooling process (Synchrotron,
ADiabatic, iC-scatterings) and on how much electrons cool during the magnetic field life-time $\tB$. For electron cooling
dominated by radiative processes (SY, iC), the timescale $\tB$ sets how long electrons cool and radiate. For AD electron-cooling,
the timescale $\tB$ determines only how long electrons radiate; they cool after $\tB$ but that is irrelevant if no emission 
is produced.

 In addition to the dominant electron-cooling process, the energy distribution of the cooling GRB electrons (the cooling-tail) 
that sets the GRB low-energy spectral slope $\bLE$ also depends on the history of the electron injection rate $R_i$ and
of the magnetic field $B$. Furthermore, $R_i(t)$ and $B(t)$ also determine the GRB pulse duration and shape. The initial
assumption was that both quantities are constant until a certain time, $\tB$ and $\ti$, respectively. This simplification
does not change much the ability of radiative processes with a cooling-power $P(\gamma) \sim \gamma^n$ of exponent 
$n \geq 2$ to account for the GRB low-energy slope $\bLE$, but a variable injection rate $R_i(t)$ is essential for allowing 
the AD-dominated electron cooling to account for more than two values for the slope $\bLE$ (1/3 and -3/4) and for 
iC-dominated cooling through scatterings at the T-KN transition of the synchrotron photons of energy below the GRB peak-energy 
$\Ep$ ($n \leq 1$) to accommodate GRB low-energy slopes softer than $\bLE = 0$. 

{\sl Hardest low-energy slope}.
 If GRB electrons do not cool well below their initial energy $\gi$ or do not radiate SY emission while they cool below 
$\gi$ (either being due to a magnetic field life-time $\tB$ shorter than the initial electron-cooling timescale $\tci$), 
the resulting slope $\bLE = 1/3$ of the instantaneous spectrum is the hardest that SY emission (not self-absorbed, sic!) 
can produce, which is a trivial fact. 

{\sl Intermediate low-energy slope}.
 Longer-lived magnetic fields yield softer slopes $\bLE$ for the integrated spectrum, with an anti-correlation between 
life-time $\tB$ and slope $\bLE$ (longer life-times lead to softer slopes) existing for $\tB \in (1,10) \tci = (1/3,3)\tgx$, 
where $\tgx \simeq 3\,\tci$ is the transit-time for electrons to migrate from emitting SY radiation at $\Ep \simeq 100$ keV
(the GRB peak-energy) to 10 keV.

{\sl Softest low-energy slope}.
 For longer magnetic field life-times $\tB > 10\, \tci$, the slope $\bLE$ of the {\sl instantaneous} spectrum settles at an
asymptotic value that depends on the dominant electron-cooling process: for radiative cooling with a cooling power exponent
$n$, the resulting slope is $\bLE = -(n-1)/2$ (for SY-cooling with $n=2$, the slope $\bLE=-1/2$ is a textbook result),
for AD-cooling, $\bLE = 0.75(1-y)$ with $y$ the exponent of the power-law electron injection rate $R_i \sim t^{-y}$,
provided that $5/9 < y < 2$ ($\bLE = 1/3$ for $y < 5/9$ and $\bLE = -3/4$ for $y > 2$). 

{\sl Pulse-integrated spectrum}.
 If electron injection lasts $\ti > \tgx$, then the pulse-{\sl integrated} spectrum has the same slope as the instantaneous 
spectrum for all radiative processes (another trivial fact), with a possible change from a cooling-tail with $n=2/3$ to one 
with $n=1$ for iC-cooling dominated by scatterings at the T-KN transition. For AD-cooling, the crossing of the lowest or 
highest-energy electrons in the cooling-tail below the observing energy leads to a soft slope $\bLE = -3/4$.

 If the electron injection lasts $\ti < \tci$ then, for radiative processes with $n \geq 2$, the cooling-tail width shrinks
after the end of electron injection at $\ti$ and the passage of the quasi-monochromatic cooling-tail below the observing 
energy leads to a GRB pulse-integrated spectrum with the same low-energy spectral slope $\bLE = -(n-1)/2$ as for a long
lived electron injection. For iC-scatterings at the T-KN transition ($n \leq 1$) and AD-cooling, the cooling-tail width
increases or remains constant, respectively, in log(energy), and the previous results for the integrated spectrum for a
longer-lived electron injection remain unchanged.

 Summarizing the above, a {\sl magnetic field life-time $\tB \in (1,10)\, \tci = (1/3,3)\,\tgx$ maps the GRB low-energy slope}
$\bLE \in [-1/2,1/3]$ if SY-cooling is dominant, 
$\bLE \in [-(n-1)/2,1/3]$ if iC-cooling in Thomson regime is dominant,
$\bLE \in [0,1/3]$ if iC-cooling at T-KN transition is dominant, and 
$\bLE \in [-3/4,1/3]$ if AD-cooling is dominant,
with the softest values for the first two cooling processes applying to short-lived ($\ti < \tci$) electron injections, 
and all softest values applying to long-lived ($\ti > 10\,\tci$) injections. 

 The measured distribution $P(\bLE)$ for the low-energy slopes of the pulse-integrated spectra does not have peaks at the 
above extreme values: the hard $\bLE = 1/3$ and the soft $\bLE = -1/2,-3/4$. That discrepancy is alleviated in part by 
the typically reported statistical uncertainty $\sigma (\bLE) \simeq 0.1$ in measuring the low-energy slope.
 Still, it is unlikely that spreading a multi-modal distribution of low-energy slopes with a kernel of dispersion 0.1
could lead to a smooth distribution $P(\bLE)$ peaking at $\bLE = 0$, particularly on its soft side with $\bLE < 0$, 
displaying the largest gap being between the preferred values $\bLE = 0$ and $\bLE = -1/2$. Thus, absent some more 
substantial systematic errors in measuring the low-energy slope, the observed quasi-Gaussian $P(\bLE)$ distribution 
requires that the distribution of magnetic field life-times $P(\tB)$ among GRB pulses is restricted to mostly 
$\tB \in (1,5)\, \tci$ and peaks at $\tB \simeq 3\,\tci$ (which yields the peak of $P(\bLE)$ at $\bLE \simeq 0$), 
without a substantial fractions of pulses with $\tB < \tci$ or $\tB > 5\, \tci$.

 At this point, such a {\sl correlation} between the magnetic field life-time $\tB$ and the cooling timescale of GRB 
electrons $\tci$ is unwarranted and puzzling.

 The conclusion that intermediate GRB slopes $\bLE \sim 0$ require an incomplete electron cooling (meaning that electrons
cool for a time $\tB$ that ranges from less than one cooling timescale $\tci$ of the typical GRB electron to at most ten $\tci$), 
is also suggested by the work of Kumar \& McMahon (2008), who analyzed the 5-dimensional model parameter space for the 
hard $\bLE = 1/3$ and soft $\bLE = -1/2$ GRB low-energy slopes, but considering that the electron cooling stops after a
re-acceleration timescale, which has the same effect on electron cooling as the disappearance of the magnetic field used 
here. At first sight, none of the possible mechanism for particle acceleration and magnetic field generation (at shocks, 
by instabilities, through magnetic reconnection) offers a reason for a correlation between that partial electron cooling 
on a timescale $\tB$ and the electron cooling timescale $\tci$. 





\newpage 

\appendix


\section{\bf A. Spectra and Light-Curves of Synchrotron Emission from Inverse-Compton (iC) Cooling Electrons}

 Inverse-Compton cooling comes in two flavors: \\ 
1) {\sl strong/fast} cooling with an exponent $n > 1$, similar to SY-dominated cooling, where 
 1a) the electron energy decreases like a power-law in time, 
 1b) the cooled-electrons distribution shrinks to quasi mono-energetic after electron injection ends 
  (the higher the exponent $n$, the faster the cooling-tail shrinks), and 
 1c) the passage of the peak-energy of the SY spectrum of the cooling-tail through the observing band 
  softens the integrated spectrum, \\
2) {\sl weak/slow} cooling with $n < 1$, 
  similar to AD-dominated cooling ($n=1$) in that the cooling-tail persists after the end of electron injection 
  and the width (in log space) of that cooling-tail is practically constant, but different from an AD cooling-tail
  in that 
 2a) the electron cooling is slower than a power-law in time, and 
 2b) the passage of the cooling-tail through the observing band does not lead to an integrated spectrum significantly 
 softer than the instantaneous spectrum.

\vspace{2mm}
\subsection{\bf A1. Strong iC-Cooling ($n > 1$) through Thomson Scatterings of High-Energy Photons} 

 This case is a generalization of SY-dominated cooling ($n=2$) and is relevant for inverse-Compton cooling when
the $\gi$-electrons scatter their own SY photons in the Thomson regime ($n \geq 2$). 
 Integrating the iC cooling law of Equation (\ref{iCcool}), one obtains
\begin{equation} 
 \gm (t) = \gi \left( 1 + \frac{t}{\tici} \right)^{-1/(n-1)} , \; \tici \equiv \frac{\gi m_ec^2}{P_{ic}(\gi)} \sim \gi^{1-n}
\label{gm}
\end{equation}
if the iC-cooling timescale $\tici$ of the $\gi$ electrons is time-independent. 

 After the end of electron injection ($t > \ti > \tici$), Equation (\ref{gm}) shows that the bounds of the cooling 
tail, $\gm(t)$ for the electrons injected initially and $\gamma_M = \gm(t-\ti)$ for the electrons injected at $\ti$,
and its width evolutions are
\begin{equation} 
 \frac {\gamma_M}{\gm} \equiv \frac{\gm(t-\ti)}{\gm(t)} = \left( 1 - \frac{\ti}{\tici+t} \right)^{-1/(n-1)} \simeq 
    1 + \frac{1}{n-1} \frac{\ti}{t} \quad {\rm if} \; t \gg \ti (> \tici) \; \ra \;
 \frac{\Delta \gamma}{\gm} = \frac{\gamma_M -\gm}{\gm} \sim \frac{\ti}{t} \ll 1 
\end{equation}
meaning that the cooling-tail becomes quasi-monoenergetic. 

 The SY peak-energy for the lowest-energy electrons $\gm$ and the transit-time from GRB peak-energy to an observing 
energy $\eps$ are
\begin{equation}
 \em (t > \tici) \simeq \Ep \left(\frac{\tici}{t} \right)^{2/(n-1)} ,\; \teic \simeq \tici \epstoEp^{-(n-1)/2}
\label{teic1}
\end{equation}
while the SY peak flux at $\em$ is 
\begin{equation}
 \fp (t) = \left\{ \begin{array}{lll} 
    \Fp(\ti) (t/\ti)                                     &  t < \ti < \tici          \\ 
    \fp(\ti) = \Fp(\ti)                                  &  \ti < t  \;  (\ti < \tici) \\ 
    \Fp(\tici) (t/\tici)                                   &  t < \tici < \ti          \\ 
    \Fp(\tici) (\em/\Ep)^{-(n-1)/2} = \Fp(\tici) (t/\tici)  &  \tici < t < \ti          \\ 
    \fp(\ti) = \Fp(\tici) (\ti/\tici)                      &  \tici < \ti < t
    \end{array} \right.
\end{equation}
with $\Fp(\ti)$ and $\Fp(\tici)$ being the GRB peak flux (the energy density at the spectral peak or at the pulse peak).
The first and third branches show the linear increase of the numer of electrons radiating at the GRB peak-energy $\Ep$
(for a constant electron injection rate), the second and fifth branches arise from the constant number of electrons 
radiating at the peak-energy $\em$ after the end of electron injection, and the fourth branch arises from the linear
increase of the number of electrons radiating at $\em$, with the flux being independent of the exponent $n$ of the cooling 
power, all cases assuming a constant magnetic field. 

 
 Adding the SY spectrum of Equation (\ref{sy}), but with the slope $\beta = -(n-1)/2$ above the peak-energy $\em$, 
and the larger-angle emission emerging at $t > \te, \te + \ti$ (owing to the exponential cut-off of the synchrotron 
function and to the quasi-monoenergetic cooling-tail), the resulting {\bf instantaneous} spectrum and {\bf pulse light-curve} 
at an observing energy $\eps < \Ep$ (below the GRB spectrum peak-energy) are
\begin{equation}
 (\ti < \tici) \quad \fe (t) \simeq \Fp(\ti) \times \left\{ \begin{array}{lll} 
   \hh   \epstoEp^{1/3} \left\{ \begin{array}{lll} 
   \hh    t/\ti                   & t < \ti         &  \quad (rise) \\ 
   \hh    \left[1+(t/\tici)\right]^{2/(3n-3)} & \ti < t < \tici  &  \quad (very\; slow \; rise) \\ 
   \hh    (t/\tici)^{2/(3n-3)}     & \tici < t < \te  &  \quad (slow \; rise) 
                          \end{array} \right.   \\
       1             & \hspace*{-58mm}  t=\te & \hspace*{14mm} (peak) \\
   \hh   \epstoEp^{-5(n-1)/6} \left( \ds \frac{t}{\tici} \right)^{-5/3} & \hspace*{-58mm} \te < t &  \hspace*{14mm} (LAE-fall) \\
                                   \end{array} \right. 
\label{feic0}
\end{equation}
\begin{equation} 
 (\tici < \ti < \te) \quad \fe (t) \simeq \Fp(\tici) \times \left\{ \begin{array}{lll} 
  \hh   \epstoEp^{1/3} \left\{ \begin{array}{lll} 
    \hh        t/\tici                         & t < \tici           &  \quad (rise) \\ 
    \hh        (t/\tici)^{(n-1/3)/(n-1)}       & \tici < t < \ti     &  \quad (fast \; rise) \\ 
    \hh       (\ti/\tici) (t/\tici)^{2/(3n-3)} & \ti < t < \te       &  \quad (slow \; rise) 
                          \end{array} \right.   \\
    \ti/\tici     & \hspace*{-49mm} t=\te  & \hspace*{12mm} (peak) \\
 \hh  \epstoEp^{-5(n-1)/6} \ds \frac{\ti}{\tici} \left( \frac{t}{\tici} \right)^{-5/3} & \hspace*{-49mm} \te < t & 
           \hspace*{12mm} (LAE-fall) \\
                       \end{array} \right. 
\label{feic1}
\end{equation}
\begin{equation} 
 (\tici < \te < \ti) \quad \fe (t) \simeq \Fp(\tici) \times \left\{ \begin{array}{lll} 
  \hh    \epstoEp^{1/3} \left\{ \begin{array}{lll} 
    \hh       t/\tici                         & \hspace*{6mm} t < \tici        &  \hspace*{6mm} (rise) \\ 
    \hh       (t/\tici)^{(n-1/3)/(n-1)}       & \hspace*{6mm} \tici < t < \te  &  \hspace*{6mm} (fast \; rise) \\ 
                          \end{array} \right.   \\
  \hh \epstoEp^{\h -(n-1)/2} \h \left\{ \begin{array}{lll} 
             1                                           & \hspace*{8mm}\te < t < \ti + \te   & (top \;plateau) \\   
        \hh  \left( \ds \frac{t}{\ti} \right)^{-(n+3)/2} & \hspace*{8mm} \ti + \te < t  & (LAE-fall)
                                   \end{array} \right.   \\ \end{array} \right. 
\label{feic2}
\end{equation}

 Thus, the pulse-peak flux and epoch are
\begin{equation}
 \tpk = \left\{ \begin{array}{lll} \te & \ti < \te \\ \te -- (\te + \ti) & \te < \ti  \end{array} \right. , \quad
 f_{pk} = \fe (\tpk) = \left\{ \begin{array}{lll} 
     \hh   \Fp(\ti) & \ti < \tici \; (<\te) \\ 
     \hh   \Fp(\tici) (\ti/\tici) & \tici < \ti < \te \\ 
     \hh   \Fp(\tici) (\eps/\Ep)^{-(n-1)/2} & \te < \ti  
           \end{array} \right. 
\label{fptpic}
\end{equation}

 The LAE flux above was calculated by assuming that its asymptotic flux decay $\fe^{(LAE)} \sim t^{-2+\beta}$ 
is continuous at the pulse peak of Equation (\ref{fptpic}) 
\begin{equation}
  \fe^{(LAE)} (t > \tpk) =  f_{pk} \times \left\{ \begin{array}{ll} 
              \hh     (t/\tpk)^{-2 + 1/3} & \ti < \te \\ 
              \hh     (t/\tpk)^{-2-(n-1)/2} & \te < \ti \end{array} \right.
\end{equation}

 From Equations (\ref{feic0}) and (\ref{feic1}), for a sufficiently short electron injection ($\ti < \te$) or a 
sufficiently low observing energy $\eps$, the SY spectrum {\bf integrated} until after the pulse peak-time $\te$ is 
\begin{displaymath}
  \Fe (t > \te > \ti) \simeq \tpk f_{pk} \simeq 
      \te \left[ \epstoEp^{1/3} \hh \left( \frac{\te}{\tici} \right)^{2/(3n-3)} \hh = 1 \right] \times
      \left\{  \begin{array}{ll}  \hh \Fp(\ti) & \ti < \tici \;(<\te) \\ 
                                  \hh \Fp(\tici) (\ti/\tici) & \tici < \ti < \te  
         \end{array} \right.
\end{displaymath}
\begin{equation}
     = \epstoEp^{-(n-1)/2} \times  
        \left\{  \begin{array}{lll} \hh \tici \Fp(\ti) & \ti < \tici \;(<\te) \\
                                    \hh  \ti \Fp(\tici) & \tici < \ti < \te  
      \end{array} \right.
\end{equation}
after using Equation (\ref{teic1}), with the pre pulse-peak and post pulse-peak (LAE) fluxes having comparable contributions 
to the pulse fluence. 

 Thus, for $\ti < \te$, the addition of instantaneous $\fe \sim \eps^{1/3}$ {\sl hard} spectra until the transit-time $\te$,
 when the typical energy $\em$ of the SY emission from the quasi-monoenergetic cooling-tail crosses 
the observing energy $\eps$, leads to a much {\sl softer} integrated spectrum $\Fe (t < \te) \sim \eps^{-(n-1)/2}$. 
 Furthermore, it can be shown that the addition of instantaneous $\fe \sim \eps^{-5(n-1)/6}$ LAE {\sl soft} spectra after 
the transit-time $\te$ leads to a {\sl harder} integrated spectrum with the same spectral slope $-(n-1)/2$.

 Thus, the passage of the peak-energy $\em$ of the cooling-tail SY emission softens the contribution of the pre pulse-peak 
emission to the integrated spectrum and hardens the contribution of the post pulse-peak emission, bringing them to the same 
integrated spectrum $\Fe (t > \tpk) \sim \eps^{-(n-1)/2}$, which is the slope of the cooling-tail SY instantaneous spectrum 
while electrons were injected at $t < \ti$. 
 This coincidence, which is also obtained for a weaker cooling process of exponent $n < 1$ (next section), arises from  
the correlation between the evolution of the peak-energy $\em$ (Equation \ref{gmic}) and the SY spectral slope $\beta (>\em)$, 
both of which are set by the exponent $n$ of the electron-cooling law.

 In the case of a sufficiently long electron injection ($\ti > \te$) or a sufficiently high observing energy $\eps$,
Equation (\ref{feic2}) gives for the SY spectrum integrated until after the pulse plateau 
\begin{equation}
 \Fe (t > \ti > \te) \simeq \tpk f_{pk} = \ti \Fp(\tici) \epstoEp^{-(n-1)/2} 
\end{equation}
as for the above $\ti < \te$ case.
 The integrated spectrum has the same slope as the instaneous spectrum because the former is dominated by the emission 
at the pulse plateau ($\te,\te +\ti$), whose duration $\ti$ is independent of the observing energy $\eps$, so that 
the plateau flux $f_{pk}$ (third branch of Equation \ref{fptpic}) imprints its energy dependence on the integrated spectrum.


 The initial assumption of a constant cooling timescale $\tici$ for the typical GRB electron of energy $\gi$ allows all 
the results for SY-dominated electron cooling shown in \S\ref{SY} to be recovered by setting the cooling power exponent $n=2$.

\vspace{2mm}
\subsection{\bf A2. Weak iC-Cooling ($n < 1$) through Scatterings of Low-Energy Photons at the Thomson--Klein-Nishina Transition} 

 The only case with $n < 1$ is that of the electron-cooling dominated by iC-scatterings if the $\gi$ electrons scatter
their SY photons in the KN regime, and at times before the iC-cooling timescale $\tici$ of the $\gi$ electrons.
In this case, the iC-cooling power $P_{ic}(\gamma) \sim \gamma^n$ has an exponent $n \simeq 2/3$. 

 From the iC-cooling law of Equation (\ref{iCcool}), the lowest electron energy $\gm$, its SY photon energy $\em$, 
and the transit-time $\te$ from emission at gamma-ray energy $\Ep$ to the observing energy $\eps$ are:
\begin{equation}
 \gm (t < \tici) = \gi \left( 1-\frac{t}{\tici} \right)^3 ,\; \tici \equiv  \frac{3\gi m_ec^2}{P_{ic}(\gi)}
\label{gmic}
\end{equation}
\begin{equation}
 \em (t < \tici) = \Ep \left( 1-\frac{t}{\tici} \right)^6 ,\; \teic = \tici \left[1-\epstoEp^{1/6} \right] 
\label{teic2}
\end{equation}
assuming a constant magnetic field $B$ and a constant iC-cooling timescale $\tici \equiv \tic(\gi)$. 

 In constrast with the $n>1$ case, for $n < 1$, the observing energy $\eps$ is "reached" before the cooling timescale 
$\tici$ of the $\gi$ electrons, and the spectrum $\fe \sim \eps^{1/6}$ of the SY emission from the cooled electron 
distribution rises (instead of falling), peaking at $\ep = \Ep$, the GRB energy peak. Thus, the flux at $\em$ is
\begin{equation}
 \fm (t < \ti) = \Fp(t) \left( \frac{\em}{\Ep} \right)^{1/6}, \; \Fp (t<\ti) = \Fp(\ti) \frac{t}{\ti} \quad
   \ra \quad \fm (t < \ti) = \Fp(\ti) \frac{t}{\ti} \left(1 - \frac{t}{\tici} \right) 
\end{equation}
where the evolution of the GRB peak flux $\Fp (t<\ti)$ stands for a constant electron injection rate $R_i$
and a constant magnetic field $B$.

 The $\gi$-electrons injected at $\ti$ cool following
\begin{equation}
 \gamma_p (t > \ti) = \gm (t-\ti) =  \gi \left( 1-\frac{t-\ti}{\tici} \right)^3 \ra  \;
 \frac{\gp(t)}{\gm(t)} = \left( 1+\frac{\ti}{\tici-t} \right)^3 
\end{equation}
using Equation (\ref{gmic}), thus the width of the cooling-tail increases slowly at $\ti < t \ll \tici$, implying that 
the cooling-tail is stretched and becomes harder.
That hardening being slow, we will ignore it and assume that the cooling-tail remains a power-law of exponent $-n$.
 The SY spectrum peaks at the characteristic SY energy of the $\gp$-electrons, $\ep = \esy(\gp)$, 
and the peak flux at that energy is approximately constant (if $B=const$) 
\begin{equation}
 \fp (t > \ti) = \Fp(\ti) \;,\; \ep (t) = \Ep \left( \frac{\gp}{\gi} \right)^2
\label{fpep}
\end{equation}
because most electrons (of constant number) radiate at $\ep$.

 From the SY spectrum corresponding to the broken power-law electron distribution 
\begin{equation}
 \fe (t) = \left\{ \begin{array}{lll}  
    \hh \fm (\eps/\em)^{1/3} & (\eps < \em) & (t < \te) \\ 
    \hh \fm (\eps/\em)^{1/6} = \fp (\eps/\ep)^{1/6} & (\em < \eps < \ep) & (\te < t < \ti + \te) \\ 
    \hh \fp (\eps/\ep)^{-(p-1)/2} & (\ep < \eps) & (\ti + \te < t) 
   \end{array} \right.
\end{equation}
and the above evolutions of break energies $\em$, $\ep$ and of the coresponding fluxes $\fm$ and $\fp$, one can
calculate the {\bf instantaneous} spectrum and pulse {\bf light-curve} at observing energy $\eps$ 
\begin{equation}
 (\ti < \te) \quad \fe (t) = \Fp(\ti)  \left\{ \begin{array}{lll} 
   \hh \epstoEp^{1/3}  \left( 1- \ds \frac{t}{\tici} \right)^{-1} \left\{ \begin{array}{lll} 
          \ds \frac{t}{\ti}                               & t < \ti           & (fast \; rise) \\ 
     \hh  \left( 1- \ds \frac{t-\ti}{\tici} \right)^{-1}   & \ti < t < \te     & (rise) 
                \end{array} \right. \\
  \hh \epstoEp^{1/6} \left(1-\ds\frac{t-\ti}{\tici}\right)^{-1} & \hspace*{-50mm} \te<t<\te+\ti & (slow \; rise) \\
     1       & \hspace*{-50mm} \te + \ti    & (peak)   \\
  \hh \epstoEp^{-(p-1)/2}  \left( 1- \ds \frac{t - \ti}{\tici} \right)^{3(p-1)} &  \hspace*{-50mm} \ti + \te < t  & (fall) 
   \end{array} \right.
\label{fenless1}
\end{equation}

\begin{equation}
 (\te < \ti < \tici) \quad \fe (t) = \Fp(\ti)  \left\{ \begin{array}{lll} 
 \hh \epstoEp^{1/3} \ds \frac{t}{\ti} \left( 1-\frac{t}{\tici} \right)^{-1} & \hspace*{-50mm} t < \te & \hspace*{7mm} (fast \; rise) \\ 
  \hh \epstoEp^{1/6} \left\{ \begin{array}{lll} 
        t/\ti              & \hspace*{10mm} \te < t < \ti  & (rise) \\ 
        \ds \left( 1- \frac{t-\ti}{\tici}  \right)^{-1}   & \hspace*{10mm} \ti < t < \ti +\te & (slow \; rise)  
                \end{array} \right. \\
         1         &  \hspace*{-50mm} t = \te + \ti & \hspace{7mm} (peak) \\
 \hh \epstoEp^{-(p-1)/2}  \left( 1- \ds \frac{t - \ti}{\tici} \right)^{3(p-1)} & \hspace*{-50mm} \ti + \te < t & \hspace*{8mm} (fall) 
       \end{array} \right.
\label{fenless2}
\end{equation}
with the flux at $t > \te + \ti$ as given on the last line of Equation (\ref{fenless1}).

 This shows that pulse peak-epoch and peak-flux are
\begin{equation}
  \tpk = \te +\ti \;,\quad  f_{pk} = \fe (\tpk) = \Fp(\ti) 
\label{fptpic1}
\end{equation}
where $\Fp(\ti)$ is the GRB pulse peak-flux (or the GRB peak spectral energy). The peak epoch $\tpk$ corresponds
to the passage of the high-energy end $\ep$ of the cooling-tail, after the end of electron injection.

 Equations (\ref{fenless1}) and (\ref{fenless2}) show that the SY spectrum integrated until the transit-time $\te$ 
(when the lowest energy $\em$ of the power-law SY spectrum from the cooling-tail crosses the observing energy $\eps$) 
is $\Fe \sim \eps^{1/3}$ and that, after $\te$, it is $\Fe \sim \eps^{1/6}$, with the $\ep$ crossing at the peak-time 
$\tpk = \te + \ti$ yielding a contribution with the same spectral slope 1/6: $\Fe (t > \tpk) - \Fe(t=\tpk) \sim 
\eps^{-(p-1)/2} (1-\te/\tici)^{3p-2} \sim \eps^{1/6}$, after using Equation (\ref{teic2}). 

 The same two equations show that the slow pre-peak rise and post-peak fall after $\max (\te,\ti)$ are always dominant 
over the preceding fluence, the integrated flux being 
\begin{equation}
  \Fe (t \gg \tpk) \simg \Fp (\ti) \tici \epstoEp^{1/6} \left\{ \begin{array}{lll} 
  \ds \hh \frac{1}{6} \ln \frac{\Ep}{\eps} + \frac{1}{3p-2} & (\te < \ti < \tici \;, \; \tilde{\eps} < \eps) \\
  \ds \hh \ln \left[1+ \frac{\ti}{\tici} \left(\frac{\Ep}{\eps}\right)^{1/6} \right]+\frac{1}{3p-2} & (\ti < \te < \tici \;, \; 
      \eps < \tilde{\eps}) \end{array} \right. \quad \tilde{\eps} \equiv \Ep \left( 1 - \frac{\ti}{\tici} \right)^6
\label{intFic}
\end{equation}
with the last term representing the contribution from the pulse fall, which can be dominant over the pre-peak contribution 
depending on the observing energy $\eps$ and on the index $p$ of the injected electron distribution. 

 The above derivations pertain to the case when the lowest-energy $\gm$ electrons in the cooling-tail cool mostly by scattering 
SY photons of energy $m_e c^2/\gm < \em$ at the T-KN limit. Those photons have a $\fe \sim \eps^{1/3}$ distribution with energy, 
leading to $P_{ic}(\gamma) \sim \gamma^{2/3}$ and to a cooling-tail distribution $\calN (\gm < \gamma < \gi) \sim \gamma^{-2/3}$,
with $\gm$ given in Equation (\ref{gmic}) for $n=2/3$. The iC-cooling power of the $\gm$ electrons switches exponent from $n=2/3$ 
to $n=1$ when the $\gm$ electrons scatter their own SY photons of energy $e_{sy}(\gm)$ at the T-KN transition (P19), i.e. when 
$\Ep' (\gm/\gi)^2 = m_e c^2/\gm$, with $\Ep'$ the GRB spectral peak-energy in the comoving frame. After that epoch, a softer
distribution $\calN (\gm < \gamma) \sim \gamma^{-1}$ grows above the low-energy end of the cooling-tail, up to an electron energy
that increases in time, i.e. the harder distribution $\calN (\gamma < \gi) \sim \gamma^{-2/3}$ of the cooling-tail below the
high-energy end $\gi$ shrinks progressively. When the $\gi$ electrons scatter the lowest energy $\em$ SY photons at the T-KN 
transition, i.e. when $\em = m_e c^2/\gi$, the entire cooling-tail becomes $\calN (\gm < \gamma < \gi) \sim \gamma^{-1}$. 
Adding that, 
for $n=1$, the cooling of the $\gm$ electrons is a exponential in time (with timescale $\tici$) that continues after the 
modified power-law cooling given in Equation (\ref{gmic}), it can be shown that the $n=2/3$ initial iC cooling-tail is 
completely replaced by a softer $n=1$ cooling-tail at epoch
\begin{equation}
 \ttic = \tici \left[ 1 - \left( \frac{m_e c^2}{\gi \Ep'} \right)^{1/9} + \frac{1}{6} \ln \frac {\gi \Ep'}{m_e c^2} \right] 
\label{ttic}
\end{equation}

 For the $n=1$ cooling-tail to develop up to the GRB typical electron energy $\gi$, electron injection must last longer
than the iC switch-time $\ttic$: $\ti > \ttic$ (first condition).
For the $n=1$ cooling-tail SY emission to dominate the integrated spectrum, the iC switch-time $\ttic$ must occur before 
the pulse-peak epoch $\tpk = \ti + \te$, with the transit-time $\te$ for iC-cooling with $n=2/3$: $\ti > \ttic - \te$
(second condition). The second condition is satisfied if the first one is fullfilled, thus, the $n=1$ cooling-tail 
yields an integrated spectrum of slope $\bLE = 0$ only if $\ti > \ttic$.

\vspace{2mm}
\subsection{\bf A3. Pulse Duration and Transit-Time}

\vspace*{1mm}
\subsubsection{A3.1 \hspace{2mm} \sl Constant $B$ and Increasing $\tau(t)$ (Decreasing $\tic(\gi)$)}

 To assess the robustness of the above result regarding GRBs with a hard low-energy slope $\bLE > 0$ arising from 
iC-dominated electron cooling with $n < 1$, we consider next the case when the scattering optical-thickness $\tau$ 
is not constant. 
 For an electron injection rate $R_i \sim t^y$, the above case of a constant $\tau$ (leading to a constant iC-cooling 
timescale $\tic (\gi)$) corresponds to $y < -1$. 
 For $y > -1$, when $\tau \sim t^{y+1}$ increases, the cooling timescale $\tici \sim \tau^{-1}$ decreases and 
the iC-cooling law of Equation (\ref{iCcool}) becomes
\begin{equation}
 - \frac{d\gamma}{dt} = \frac{1}{\tici(\ti)} \left( \frac{t}{\ti} \right)^{y+1} \frac{\gamma^n}{\gi^{n-1}} \quad (y \geq -1)
\label{iccool}
\end{equation}
with the cooling timescale $\tici(\ti)$ at the end of electron injection containing all relevant and unspecified 
quantities: magnetic field $B$ and optical-thickness $\tau (\ti)$.

 The above equation can be integrated to derive $\gamma(t)$ and, by using $\eps/\Ep = [\gamma(\te)/\gi]^2$ as
definition for the transit-time $\te$, one obtains
\begin{equation}
 \teic = \ti \left\{ \frac{y+2}{n-1} \left[\epstoEp^{(1-n)/2} -1 \right] \frac{\tici}{\ti} \right\}^{1/(y+2)} \; (y \geq -1)
\label{teicy}
\end{equation}
The pulse duration can be calculated as in Equation (\ref{dtic}):
\begin{equation}
  \dteps =  \epstoEp^{(1-n)/2} \left( \frac{\ti}{\te} \right)^{y+1} \tici(\ti) \quad (y \geq -1)
\label{dtic1}
\end{equation}
For $y = -1$, Equations (\ref{teicy}) and (\ref{dtic1}) give timescales for a constant scattering optical-thickness 
$\tau$.  For an increasing $\tau(t)$, the transit-time $\te$ has a weaker dependence on the observing energy $\eps$ 
than for a constant $\tau$ (Equation \ref{teic}) because the exponent $1/(y+2) < 1$, while the pulse duration $\dteps$ 
picks-up an energy-dependent factor $(\ti/\te)^{y+1}$.

 For iC-cooling dominated by scatterings in the {\bf Thomson} regime ($n > 1$), Equations (\ref{teicy}) and (\ref{dtic1}) 
lead to 
\begin{equation}
  \dteps  = \frac{n-1}{y+2} \te \sim \epstoEp^{-(n-1)/(2y+4)} (y \geq -1)
\end{equation}
thus, the trend of pulses to last shorter at higher energies still stands even when the scattering optical-thickness 
$\tau$ increases. 
The measured energy dependence of the pulse duration, $\dtg \sim \eps^{-0.4}$, has an exponent between the values 
$-0.25$ and $-0.50$ expected for iC-dominated electron cooling in the Thomson regime ($n=2$), for a constant electron 
injection rate $R_i$ ($y=0$) or a constant optical-thickness $\tau$ ($y=-1$), respectively.

 For iC-cooling dominated by scatterings at the {\bf T-KN} transition ($n < 1$), if the electron injection rate $R_i$ 
decreases sufficienly fast (faster than $1/t$), then the cooling-tail will be curved downward, with most of the flux 
being produced by the lowest energy $\gm$ electrons of the tail. Thus, the pulse peak-time will be the transit-time 
that it takes the $\gm$ electrons to cool to a SY-emitting energy equal to observing energy $\eps$, and the pulse 
duration will be set by their cooling rate when their SY characteristic energy $\em$ drops to $\eps$. 
 The exact evolution of the injection rate $R_i$ is not relevant for the electron iC-cooling because the scattering 
optical-thickness $\tau$ is practically constant. Equations (\ref{teicy}) and (\ref{dtic1}) still apply, but with $y=-1$,
for which they reduce to Equations (\ref{teic}) and (\ref{dtic}), thus, in the limit $\eps \ll \Ep$, the transit-time 
$\te$ is very weakly dependent on the observing energy $\eps$, and the pulse duration $\dteps \sim \eps^{(1-n)/2}$ 
increases with $\eps$, remaining at odds with GRB observations. For $\eps \siml \Ep$ (just below to the GRB peak-energy), 
the term $\te$ in the denominator of Equation (\ref{dtic1}) makes $\dteps$ increase with $\eps$ even stronger than 
$\eps^{(1-n/2}$, thus the incompatibility with observations becomes more severe.

 If $R_i$ does not decrease faster than $1/t$, then the cooling-tail will be close to a power-law 
or will be curved upward, with most of the flux arising from the highest energy $\gp$ electrons of the tail.
Thus, the pulse peak-epoch will be the transit-time for the $\gp$ electrons to "reach" the observing energy $\eps$
after the end of electron injection at $\ti$, and the pulse duration $\dteps$ will be set by the cooling rate of
the $\gp$ electrons when their SY characteristic $\ep$ reaches $\eps$, at $\teic > \ti$. Because the scattering
optical-thickness is constant after the end of electron injection, the iC-cooling is independent of the history 
of electron injection rate $R_i(t < \ti)$, and so are the transit-time $\te$ and pulse duration $\dteps$ arising 
from the cooling of the $\gp$ electrons after $\ti$.
Consequently, the pulse duration is still as given in Equation (\ref{dtic}), and the incompatibility with observations 
remains unchanged.

\vspace*{1mm}
\subsubsection{A3.2 \hspace*{2mm} \sl Decreasing Magnetic Field $B(t)$}
\vspace*{1mm}

 An evolving magnetic field $B(t)$ affects the transit-time $\te$ and the pulse duration $\dteps$ in two ways:
first, it determines the energy density of the seed SY photons to be upscattered and iC-cool the electrons,
thus it determines the electron cooling and, second, it determines the evolution of the ends of the SY spectrum
from the cooling-tail, whose passage through the observing band defines $\te$ and $\dteps$.

 The case of electron iC-cooling occurring mostly through scatterings in the Thomson regime ($n > 1$) will not be 
considered further because, as shown above, it can account for the observed trend of 
GRB pulse duration to decrease with observing energy, provided that electrons cool to below the observing energy, 
and we focus on iC-cooling dominated by scatterings of the sub-GRB SY photons at the {\bf T-KN} transition ($n < 1$), 
a case that fails to accommodate that observational feature for a constant magnetic field. 

 For iC-cooling with $n < 1$, the history of the electron injection rate $R_i(t)$ is irrelevant for the cooling 
of the electrons that determine the transit-time and the pulse duration, thus these two quantities depend only 
on the evolution of the magnetic field, which we will assume to be a power-law $B(t) = B_i (t/\ti)^x$, normalized 
at the end of electron injection. 
 It can be shown that the iC-cooling power satisfies $P_{ic}(\gamma) \sim \tau (B\gamma)^n$, either for $n < 1$
(equation 45 of P19) or for $n > 1$ (equation 41 of P19), thus, for the above $B(t)$, the iC-cooling law is 
\begin{equation}
 - \frac{d\gamma}{dt} = \frac{1}{\tici(\ti)} \left( \frac{t}{\ti} \right)^{nx} \frac{\gamma^n}{\gi^{n-1}}
\label{icB}
\end{equation}
which can be integrated (from $t=0$ to $t=\te$ if $\tau$ is constant and from $t=\ti$ to $t=\ti+\te$ if $\tau$
increases) and, after using
\begin{equation}
 \frac{\eps}{\Ep} = \left( \frac{\gamma(\te)}{\gi} \right)^2 \frac{B(\te)}{B_i} = 
                    \left( \frac{\gamma(\te)}{\gi} \right)^2 \left( \frac{\te}{\ti} \right)^x
\end{equation}
will lead to an algebraic equation for the transit-time $\te$, which can be solved in asymptotic regimes:
\begin{equation}
 (n < 1 ,\; B \sim t^x) \quad
 \teic \simeq \ti \times \left\{ \begin{array}{lll} 
 \hh   \epstoEp^{1/x} & \tilde{\eps} \ll \eps < \Ep & (x < 0) \\ 
 \hh   \left(\ds \frac{\tici}{\ti} \right)^{1/(nx+1)} & \eps \ll \tilde{\eps} & \hspace{-10mm} (-\frac{1}{n} < x < 0)
   \end{array} \right.   \quad
 \tilde{\eps} \simeq \Ep \left( \frac{\tici}{\ti} \right)^{x/(nx+1)}
\label{teicx}
\end{equation}
 The second branch of Equation (\ref{teicx}) shows that, for a sufficiently low observing energy $\eps$, the transit-time 
is independent of $\eps$. In the limit $x \rightarrow 0$ (constant $B$), one has $\tilde{\eps} (x=0) = \Ep$, 
thus the first branch of Equation (\ref{teicx}) disappears and the second branch reduces to the transit-time given 
in Equation (\ref{teic}) in the limit $\eps \ll \Ep$, 

 The condition $x < 0$ arises from requiring that the transit-time decreases with observing energy $\eps$; 
in the opposite case ($x > 0$), an increasing magnetic field would compensate the electron cooling and lead to
break energies (at either end of the cooling tail's SY spectrum) that increase, and there is no transit of the
break energies to an observing energy $\eps < \Ep$.
 The working condition $x > -1/n$ for the second branch leads to simple temporal power-law dependence for the 
cooling-equation solution $\gamma(t)$; for $x < -1/n$, the corresponding equation for $\te$ becomes even more 
complicated; however, the result given in the first branch stands for $x < -1/n$ as well.

 Once the transit-time $\te$ is known, the pulse duration can be calculated by using the cooling-law of Equation (\ref{icB}):
\begin{equation}
  \dteps = \epstoEp^{(1-n)/2} \left( \frac{\ti}{\te} \right)^{(n+1)x/2} \tici(\ti) 
\end{equation}
leading to
\begin{equation}
 \dteps \simeq \tici(\ti) \times \left\{ \begin{array}{lll} 
  \hh \epstoEp^{-n=-2/3} & \hh \tilde{\eps} < \eps < \Ep & (x < 0)  \\
  \hh \left( \ds \frac{\ti}{\tici} \right)^{(n+1)x/(2nx+2)} \hh \epstoEp^{(1-n)/2=1/6} & 
      \hh \eps < \tilde{\eps} & (-3/2 = -1/n < x < 0)
   \end{array} \right.   
\end{equation}
For a sufficiently low observing energy (second branch), the pulse duration still increases with energy, however, 
for energies just below the GRB spectral peak $\Ep$, a decreasing magnetic field "opens" the first branch above, 
for which $\dteps \sim \eps^{-n} \sim \eps^{-2/3}$ is consistent with (or stronger than) the observed trend of 
pulse duration to decrease with energy.

\vspace{3mm}
\section{\bf B. Spectra and Light-Curves of Synchrotron Emission from Adiabatically-Cooling Electrons}

 During electron injection (at $t < \ti$) at the power-law rate $R_i \sim (t+\to)^{-y}$, the cooling-tail extends 
from the lowest energy $\gamma_m$ given in Equation (\ref{gmad}) to the minimal electron injection energy $\gi$,
thus, the cooling-tail SY emission extends from photon energies $\em(t)$ (Equation \ref{gmad}) to 
$\ep (t < \ti) = \Ep$ and has the slope $\beta$ given in Equation (\ref{betaAD}). 
For $y > 1$, $m = (1-3y)/2 < -1$ and most cooled electrons are at $\gm$; for $y < 1$, $m > -1$ and most cooled electrons 
are at $\gi$; irrespective of $y$. 

 From $\int_{\gm}^{\gi} d\gamma \calN (\gamma) \simeq \int_0^t dt' R_i (t')$, where $\calN (\gamma) = dN/d\gamma$ 
is the electron distribution  with energy, it can be shown that
\begin{equation}
  \gm \calN (\gm) = const \ra \calN (\gm) \sim \gm^{-1} \sim t^{2/3} \;, \; 
  \gi \calN (\gi) \sim (t+\to)^{1-y} \ra \calN (\gi) \sim t^{1-y} 
\end{equation}
Thus, for a constant magnetic field $B$, the flux densities $\fe \sim B \gamma(\eps) \calN (\gamma)$ at the ends 
of the cooling-tail's characteristic synchrotron energies $\em$ and $\ep=\Ep$ are
\begin{equation}
 \fm (\em) = \Fp (\ti) \left( 1+ \frac{\ti}{\to} \right)^{y-1} = const \;, \;
 \fp (\Ep, t < \ti) = \Fp (\ti) \left( \frac{t + \to}{\ti + \to} \right)^{1-y} \simeq  
   \Fp (\ti) \left( \frac{t}{\ti} \right)^{1-y} \quad (\to \ll t,\ti)
\label{fmfp}
\end{equation}
with $\Fp(\ti)$ the flux density at the GRB peak-energy $\Ep$ when electron injection ends. For $y < 1$, i.e. for
low-energy slopes $\bLE (t < \ti) > 0$, the epoch $\ti$ also marks the GRB pulse peak, thus $\Fp(\ti)$ {\sl is 
also the GRB pulse peak-flux (or pulse-peak flux) for GRBs with a harder low-energy slope during the pulse rise}.  

 After electron injection ends (at $t > \ti$), the lowest-energy electrons $\gm$ continue to cool adiabatically
according to Equation (\ref{gmad}), thus $\em$ evolves is as in Equation (\ref{gmad}), while the $\gi$ electrons 
begin cooling adiabatically at $\ti$ following the same law but with time measured since $\ti$ and for the current 
system age $\ti +\to$:
\begin{equation}
 \gp(t)  = \gi \left( 1 + \frac{t-\ti}{\ti + \to} \right)^{-2/3} \simeq \gi \left( \frac{t}{\ti} \right)^{-2/3} 
      \ra \quad \ep (t>\ti) = \Ep  \left( \frac{t}{\ti} \right)^{-4/3} \;, \; 
    \frac{\gp(t)}{\gm(t)} = \left( \frac{\ti}{\to} \right)^{2/3}     \quad (\to \ll \ti < t)
\end{equation}
thus the width of the cooling-tail is constant.
From the electron AD-cooling law $\gamma (t>\ti) \sim t^{-2/3}$, it can be shown that the cooling-tail slope
$-m$ remains unchanged at $t > \ti$.
From $\int_{\gm}^{\gp} d\gamma \calN (\gamma) = const$ at $t>\ti$, it can be shown that the flux densities at the
at the ends $\em$ and $\ep$ of cooling-tail remain constant after $\ti$.
Thus, the flux density at the lowest energy of the cooling-tail is same as in Equation (\ref{fmfp}), while the flux 
density at the higher-energy break $\ep$ is 
\begin{equation}
 \fp(\ep,t>\ti) = \fp(\ep,\ti) \simeq \Fp(\ti) 
\label{Fp}
\end{equation}
after using Equation (\ref{fmfp}).

 Adding the SY spectrum from a broken power-law electron distribution consisting of a cooling-tail 
$\calN (\gm < \gamma < \gp) \sim \gamma^{(1-3y)/2}$ and a cooled-injected electron distribution 
$\calN (\gp < \gamma) \sim \gamma^{-p}$ is
\begin{equation}
 \fe \sim \left\{ \begin{array}{lll} 
     \fp (\eps/\ep)^{1/3}       & \eps < \ep       & (y < 5/9) \\
     \fm (\eps/\em)^{1/3}       & \eps < \em       & (y > 5/9) \\
     \fm (\eps/\em)^{3(1-y)/4}  & \em < \eps < \ep & (y > 5/9) \\
     \fp (\eps/\ep)^{-(p-1)/2}  & \ep < \eps       &  \\
   \end{array}  \right.
\end{equation}
the {\bf instantaneous} spectrum ({\bf pulse light-curve}) at lower energies can be calculated:
\begin{equation} 
 (R_i \sim t^{-y}, y < 5/9) \quad \fe(t) = \Fp(\ti) \times \left\{ \begin{array}{lll} 
  \hh \epstoEp^{1/3} \left(\ds \frac{t}{\ti} \right)^{1-y}             & t<\ti     & (rise) \\
  \hh \epstoEp^{1/3} \left(\ds \frac{t}{\ti} \right)^{4/9}             & \ti<t<\tp & (slow \; rise) \\
   1                                                                   & t = \tp   & (peak) \\
  \hh \epstoEp^{-(p-1)/2} \left( \ds \frac{t}{\ti} \right)^{-2(p-1)/3} & \tp<t     & (fall) 
   \end{array}  \right.  \quad
\label{FAD1}
\end{equation}
with $\te$ the epoch when the lowest energy electrons $\em$ radiate at the observing energy $\eps$ (Equation \ref{tead}),
and $\tp$ the epoch when the higher spectral break at $\ep$ crosses the observing energy $\eps$: 
\begin{equation} 
   \tp \equiv \ti \epstoEp^{-3/4} = \frac{\ti}{\to} \te 
\label{tpad}
\end{equation}
and
\begin{equation}
 (R_i \sim t^{-y}, 5/9 < y) \quad \fe(t) = \Fp(\ti) \left\{ \begin{array}{lll} 
 \hh  \epstoEp^{1/3} \left(\ds \frac{\ti}{\to} \right)^{y-1} \left(\ds \frac{t}{\to} \right)^{4/9} & t<\te     & (slow \; rise) \\
 \hh  \left(\ds \frac{\ti}{\to} \right)^{y-1}                                                      & t = \te   & (peak\; if\; y > 1) \\
 \hh  \epstoEp^{3(1-y)/4} \left(\ds \frac{t}{\ti} \right)^{1-y}            & \te<t<\tp & (y<1|slow\;rise ; y>1|fall) \\
   1                                                                                            & t = \tp   & (peak\; if\; y < 1) \\
 \hh  \epstoEp^{-(p-1)/2} \left( \ds \frac{t}{\ti} \right)^{-2(p-1)/3}                             & \tp<t     & (fall) 
   \end{array}  \right.
\label{FAD2}
\end{equation}
For any $y$, $\fe(\tp) = \Fp(\ti)$ follows from Equation (\ref{Fp}) because $\tp > \ti$ (Equation \ref{tpad}), 
i.e. the high-energy break $\ep$ decreases below the observing energy $\eps$ only after electron  injection stops.

 For $y < 5/9$, the cooling-tail emission is dimmer than the $\eps^{1/3}$ low-energy SY flux produced
by the $\gp$ electrons and the entire spectrum is as if the cooling-tail did not exist, thus the epoch $\te$
when $\em$ crosses the observing band is {\sl irrelevant}. 
For $y > 5/9$, the light-curve depends on the evolution of the lowest-energy characteristics ($\em$ and $\fm$), 
which are unchanged across $\ti$, thus there is no light-curve break at the epoch $\ti$ when electron injection 
stops and the epoch $\ti$ is {\sl irrelevant}.

 Equations (\ref{FAD1}) and (\ref{FAD2}) show that the pulse rise is harder ($\beta_r \geq 0$) than its decay 
($\beta_f \leq 0$), and that the pulse peak-epoch and peak-flux are 
\begin{equation}
  \tpk = \left\{ \begin{array}{ll} \tp^{(ad)} & y < 1 \\ \te^{(ad)}  & y > 1  \end{array} \right.
  =  \epstoEp^{-3/4} \left\{ \begin{array}{ll} \ti & y < 1 \\ \to & y > 1  \end{array} \right. \;, \;
  f_{pk} = \Fp(\ti) \times \left\{ \begin{array}{ll} 1 & y < 1 \\ (\ti/\to)^{y-1}  & y > 1  \end{array} \right.
\label{fptpad}
\end{equation}

 The {\bf pulse-integrated} spectrum is dominated by the emission prior to when the flux decay becomes faster 
than $t^{-1}$. That happens at $\te$ if $y > 2$ (leading to $\Fe \sim \eps^{-3/4}$, see below), at $\tp$ if $y < 2$ 
(leading to $\Fe \sim \eps^{-3/4}$, see below) and if the injected electron distribution has an exponent $p > 5/2$, 
but continues through the pulse decay at $t > \tp$ if $p < 5/2$ (leading to $\Fe \sim \eps^{-(p-1)/2}$).
After some calculations, the integrated spectrum is found to be
\begin{equation}
 (R_i \sim t^{-y}) \quad \quad \Fe(t>\tp) = \Fp(\ti) \ti \times \left\{ 
    \begin{array}{lll} \hh  \epstoEp^{-3/4} & p > 5/2 \\
                       \hh  \epstoEp^{-(p-1)/2} \left(\ds \frac{t}{\ti} \right)^{(5-2p)/3}  & p < 5/2 
   \end{array}  \right.
\label{FintAD}
\end{equation}

 The above slope $\bLE=-3/4$ of the integrated spectrum arises from the passage of the higher spectral break $\ep$, 
because the fluence from the pulse fall is most often dominant, yielding $\Fe \simeq \tp \fe(\tp) \sim 
\tp (\eps^{1/3} \tp^{4/9}) \sim \eps^{-3/4}$. However, even if when the pulse fluence is dominated by the
crossing of the lower spectral break $\em$ (i.e. for $y > 3.4$), the slope of the integrated spectrum would be
the same $\bLE = -3/4$ because the transit-times $\te$ (Equation \ref{tead}) and $\tp$ (Equation \ref{tpad}) 
have the same dependence on the observing energy $\eps$. 
 As for SY cooling, the softness of the pulse-integrated spectrum is a consequence of the cooling-time 
increase with observing energy: $\tead \sim \eps^{-3/4}$.

 The last branch above simply states that, if the injected electron distribution is sufficiently hard ($p < 5/2$), 
then the pulse fluence is mostly from the cooled-injected distribution, and not from the cooling-tail.
In this case, the integrated spectrum will have the slope $\bLE = -(p-1)/2$ of the cooled-injected distribution,
which has the slope of the injected power-law energy distribution because AD cooling shifts distributions
to lower energies while preserving their slopes.


\vspace{3mm}
\section{\bf C. Synchrotron (SY) and Adiabatic (AD) Cooling}

 Equations (\ref{sycool}) and (\ref{adcool}) lead to
\begin{equation}
 (SY+AD) \quad  -\frac{d\gamma}{dt} = -\left(\frac{d\gamma}{dt}\right)_{sy} -\left(\frac{d\gamma}{dt}\right)_{ad} = 
   \frac{\gamma^2}{\go \tsy(\go)} + \frac{2}{3} \frac{\gamma}{t+\to}   
\label{syadcool}
\end{equation}
for an electron of initial energy $\go$. In above equation, $\tsy(\go) = 7.7\ttimes 10^8/(\go B^2)$ 
(Equation \ref{gmsy}) is the SY-cooling timescale for the $\go$ electrons. With the substitution $\gamma = g^x$,
the electron-cooling equation becomes
\begin{equation}
  x g^{x-1} \frac{dg}{dt} + \frac{g^{2x}}{\go \tsy(\go)} + \frac{2}{3} \frac{g^x}{t+\to} = 0
\end{equation}
which is a first-order linear differential equation of the form
\begin{equation}
  \frac{dg}{dt} + a(t) g = b(t) \quad \ra \quad g(t) = \frac{1}{\mu(t)} \left[ const + \int \mu(t) b(t) dt \right]
    \;, \quad \mu(t) \equiv \exp\left\{ \int a(t) dt \right\}
\label{linear}
\end{equation}
only if $x=-1$: 
\begin{equation}
 \gamma \equiv 1/g \; \ra \; \frac{dg}{dt} - \frac{2}{3} \frac{g}{t+\to} = \frac{1}{\go \tsy(\go)}
   \quad \ra \quad \mu(t) = (t+\to)^{-2/3} \;,\; g(t) = const(t+\to)^{2/3} + \frac{3(t+\to)}{\go \tsy(\go)}
\label{linear1}
\end{equation}
where the constant can be determined from the initial condition $\gamma(t=0) = \go$. 

 Thus, the solution to the SY and AD-cooling is : 
\begin{equation} (B=const): \quad 
   \gamma (t) =  \frac{\go}{ \ds \left( 1 + \frac{t}{\to} \right)^{2/3} \left\{ 1 + \frac{3\to}{\tsy(\go)} 
     \left[ \left( 1 + \frac{t}{\to} \right)^{1/3} - 1 \right] \right\} }
\label{syadsol}
\end{equation}
which can be written as
\begin{equation}  (AD+SY): \quad
  \gamma(t) = \frac{\gad(t)}{\ds 1 + 2\frac{X(t)-1}{Z}}  \;,\; \gad (t) = \frac{\go}{X^2} \;,\; 
    X(t) \equiv \left( 1 + \frac{t}{\to} \right)^{1/3} \;,\; Z \equiv \frac{2\tsy(\go)}{3\to} = \frac{\tsy(\go)}{\tad(t=0)}
\label{adsysol}
\end{equation}
where $\gad$ is the solution (Equation \ref{gmad}) to the AD cooling law.
Thus, {\sl the solution to the SY and AD cooling} is a {\sl modified adiabatic-cooling solution}, with denominator 
containing information about both AD and SY cooling. 
The reason for this structure for the cooling solution is that the linear term of the first-order LDE for electron cooling 
(Equation \ref{linear1}) contains only AD-cooling. 
In the limit $t \ll \to$, the denominator loses dependence on AD cooling and the solution becomes 
\begin{equation}  
 (t \ll \to): \quad \gamma(t) = \frac{\go}{\ds \left( 1+\frac{t}{\to} \right)^{2/3} \left( 1+\frac{t}{\tsy(\go)} \right)}  
    \siml \gsy (t) \equiv \frac{\go}{\ds 1 + \frac{t}{\tsy(\go)}}
\label{adsysol1}
\end{equation}
which shows SY cooling (Equation \ref{gmsy}) and AD cooling (Equation \ref{gmad}) operating independently. 

 The inverse-Compton cooling law of Equation (\ref{iCcool}) can be added to the AD and SY cooling terms of
Equation (\ref{syadcool}), but an approximative calculation of the inverse-Compton power is possible only until the overall
cooling timescale $\tci$ of the typical $\gi$ electron (i.e. before the cooling-tail develops) because, in the opposite case
$t > \tci$, the Compton parameter $Z$ depends on the minimal electron energy $\gm$ of the cooling-tail, whose evolution
is not known in advance (unless iC-cooling is weaker than SY and AD-cooling and Equation \ref{syadsol} can be used as 
an approximation of $\gm$). For $t < \tci$, the iC power depends on the electron energy as $P_{ic} \sim \gamma^{2/3}$ 
for an electron that scatter the SY photons $\Ep$ produced by the typical $\gi$-electron in the Klein-Nishina regime 
($\gamma \Ep' > m c^2$); then iC-cooling introduces a $t \gamma^{2/3}$ term in Equation (\ref{syadcool}), with time $t$
entering through the proportionality of the iC power on the electron scattering optical-thickness $\tau$. 

 In this case, the substitution $\gamma = 1/g$ that is required for the SY-cooling term of Equation (\ref{syadcool}) to yield 
a first-order linear differential equation (LDE) brings an iC-cooling term proportional to $t g^{4/3}$, which cannot be combined 
with the AD-cooling term $g/(t+\to)$ to lead to the term $a(t)g$ of the LDE (\ref{syadcool}).
 However, for electrons that scatter $\Ep$ photons in the Thomson regime ($\gamma \Ep' < m c^2$), the iC power has the same 
$P_{ic} \sim \gamma^2$ dependence as the SY power and the substitution $\gamma = 1/g$ used above leads to a cooling law of the 
form given in Equation (\ref{linear}) but with a slightly more complex term $b(t) = const + kt$.

 Then, the first integral given in Equation (\ref{linear}) can be calculated analytically and the electron cooling subject to 
all three processes is
\begin{equation}
 (AD+SY+iC/Th): \quad
  \frac{\gad}{\gamma (t)} = 1 + 2\frac{X-1}{Z} + \frac{\to^2}{\ti \tic(\go)}(X^4-4X+3)
   \;,\; \tic(\go) \equiv \frac{\tsy(\go)}{\gi^2 \tau(\ti)}
\label{fullsol}
\end{equation}
where $\tic(\go)$ is the iC-cooling timescale at the epoch $\ti$ when the electron injection ends and the scattering 
optical-thickness $\tau$ is maximal.
 

 Again, the solution (Equation \ref{fullsol}) to the full electron-cooling law (as in Equation \ref{syadcool} but with an extra 
term for iC-cooling in the Thomson regime: $-(d\gamma/dt)_{iC} = \gamma^2/[\go\tic(\go)]$) is a modified AD-cooling solution, 
with the decrease of the electron energy $\gamma$ expedited by a SY and an iC-cooling term, each expressed as their strength 
($1/\tsy$ or $1/\tic$) relative to that of AD-cooling ($1/\to$) at the beginning of electron injection. 

 Because Equation (\ref{fullsol}) has a limited applicability, as it describes electron cooling only before the cooling-tail 
develops significantly, we will not investigate it any further. Instead, we return to Equation (\ref{adsysol}) for AD 
and SY-cooling, whose asymptotic solutions can be derived in three regimes: $i)$ $t \ll \to \rightarrow X-1 \simeq 
t/(3\to) \ll 1$ leading to the SY-solution, $ii)$ $t \gg \to \rightarrow X \simeq (t/\to)^{1/3} \gg 1$ leading to the AD and
the 1/3-SY solutions, and $iii)$ $X-1 \ll Z/2$, leading to the AD-solution. 

 Depending on the relative strength of the AD and SY losses at $t=0$, quantified by $Z$, the solution given in Equation
(\ref{adsysol}) for AD+SY cooling has the following asymptotic regimes:
\begin{equation}
  (Z < 2^{4/3}-2 \simeq 0.52): \quad \gamma (t) \simeq \left\{  \begin{array}{lll} 
         \gsy         &  t \ll \to         & ({\rm early\;SY\; solution}) \\  
         \ds \frac{1}{3} \gsy  & \to \ll t & ({\rm late\;1/3-SY\; solution})   
      \end{array} \right. 
\label{exp1}
\end{equation}
\begin{equation}
  (0.52 < Z < 1): \quad \gamma (t) \simeq \left\{  \begin{array}{llll} 
         \gsy  &  t \ll \to             & ({\rm early\;SY\; solution}) \\  
         \gad  &  \to < t < \tt \in (1,\frac{19}{8})\to    & ({\rm transient\;AD\; solution}) \\
         \ds \frac{1}{3} \gsy  &  \tt \ll t & ({\rm late\;1/3-SY\; solution})  
      \end{array} \right. 
\end{equation}
\begin{equation}
  (1 < Z): \quad \gamma (t) \simeq \left\{  \begin{array}{llll} 
         \gad              &   t \ll \tt   & ({\rm early\;AD\; solution})  \\
         \ds \frac{1}{3} \gsy  & (\frac{19}{8}\to <)\; \tt \ll t & ({\rm late\;1/3-SY\; solution})   
      \end{array} \right. 
\label{exp3}
\end{equation}
where
\begin{equation}
 \tt \equiv \to \left[\left( \frac{Z}{2}+1 \right)^3-1 \right] 
\label{tt}
\end{equation}
is the AD-SY solution switch-time, when $X-1 = Z/2$.
The above electron cooling through the three asymptotic regimes is indicated in {\bf Figure C4} by horizontal arrows
(increasing time toward right). Which asymptotic solutions are encountered depends on the parameter $Z$. 
The AD and SY solutions are separated by the $Z=2(X-1)$ line corresponding to $t=\tt$.

 The above asymptotic solutions have some interesting features: 

$i)$ Electron cooling is {\sl asymptotically} described at {\sl early times} by the {\sl SYnchrotron solution $\gsy$ 
  only if $Z < 1$}, which means $\tsy(\go) < \tad (t=0)$, i.e. only if the electron cooling is initially SY-dominated
  (obviously). Furthermore, the SY-cooling solution is accurate only at times $t \ll \to$, 
  when $X-1 \simeq t/3 \to$, for which the right-hand side term of Equation (\ref{syadsol}) shows an AD-cooling 
  term smaller than the SY-cooling term, owing to that $Z < 1$. 

$ii)$ Electron cooling is described by the {\sl ADiabatic solution $\gad$ only for $t \ll \tt$}, i.e. only for 
  $2(X-1) \ll Z$, when the right-hand side of Equation (\ref{adsysol}) is unity. 
  This condition is sufficient for an {\sl asymptotic AD-solution at early times if $Z > 1$}, i.e. if the electron 
  cooling is initially AD-dominated (obviously), but is not sufficient if $Z < 1$, i.e. if the electron cooling is 
  initially SY-dominated. In the latter case, competition between the adiabatic term $X^2$ and the mixed term 
  $1+2(X-1)/Z$ appearing in Equation (\ref{adsysol}) allows the AD-solution to set in at $t \simeq \to$. 

$iii)$ As can be seen in {\bf Figure C4}, irrespective of which cooling process is dominant initially, electron cooling 
 is {\sl asymptotically} described at {\sl late times} $t \gg \max(\to,\tt)$ by the {\sl 1/3-SYnchrotron solution}. 
 Condition $t \gg \to$ implies 
  $X \gg 1$, which implies that $X^2[1+2(X-1)/Z] \simeq X^2(1+2X/Z)$, and condition $t \gg \tt$ implies $X \gg Z/2$,
  which leads to $X^2(1+2X/Z) \simeq 2X^3/Z \simeq 2t/Z\to = 3t/\tsy(\go)$, thus $\gamma \simeq (1/3)\go \tsy(\go)/t
  \simeq \gsy/3$. In other words, for sufficiently late times, the product of the AD-cooling term $X^2$ and 
  the modified SY-cooling term $1+2(X-1)/Z \simeq 2X/Z$ is proportional to the "pure" SY-cooling term, leading to 
  a SY-cooling solution despite that, at late times, the electron cooling is guaranteed to be AD-dominated, 
  as shown below.

\begin{figure*}[t]
\centerline{\includegraphics[width=10cm,height=8cm]{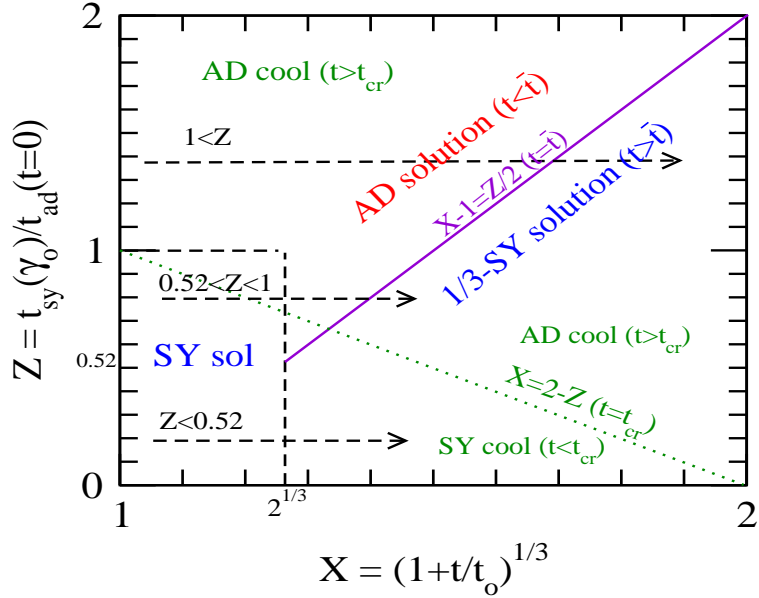}}
\figcaption{\normalsize Passage of electron cooling $\gamma(t)$ (for increasing $X(t)$) through three possible asymptotic 
 solutions (AD, SY, and 1/3-SY), depending on the initial ratio $Z$ of SY to AD cooling power ($\go = \gamma(t=0)$ is the 
 initial electron energy). (Constant magnetic field is assumed for the former). 
  The horizontal (green) arrows represent the cases identified in Equations (\ref{exp1})-(\ref{exp3}). 
 The {\bf AD-solution} ($\gad$ of Equation \ref{adsysol}) applies mostly to $Z>1$ (AD-cooling dominant initially over SY-losses) 
  and lasts until $\tt$ given by Equation (\ref{tt}) and shown by the solid (purple) line.
 The {\bf SY-solution} ($\gsy$ of Equation \ref{adsysol1}) occurs only for $Z<1$ (i.e. if SY losses are dominant over AD cooling 
  initially) and lasts until $\to$. 
 {\sl For either cooling process being dominant initially} (i.e. for any $Z$), {\sl the electron cooling eventually turns to the}
 {\bf 1/3-SY solution}, at a time $\max\{\to,\tt\}$, and that transition occurs even when the electron cooling is AD-dominated 
 after that time.
 Dotted (black) line shows the epoch when AD and SY cooling powers are equal, at $t=\tcr$ (Equation \ref{tgcr}), 
 but that epoch is irrelevant for the electron-cooling, which is (almost) always the 1/3-SY solution across $\tcr$.
}
\end{figure*}

 From Equations (\ref{tsy}) and (\ref{adsysol}), the SY-cooling timescale of the SY+AD-cooling electron is
$\tsy (\gamma)= \tsy(\go) \go/\gamma = 1.5 Z \to \go/\gamma = 3t + \tsy(\go) (t/\to)^{2/3}$ at $t \gg \to$,
while the AD-cooling timescale (Equation \ref{tad}) $\tad = 1.5 (t + \to)$ is independent of the electron energy. 
That $\tsy > 3t$ and $\tad \simg 1.5t$ guarantees that, after some time, $\tsy > \tad$ and the electron cooling 
will be eventually AD-dominated even if it started in a SY-dominated regime ($\tsy (\go) < \tad(\to)$).
The condition $\tsy (\gamma) = \tad$ implies $X=2-Z$, which defines a critical time and a critical electron energy
\begin{equation}
 (Z < 1): \quad 
 \tcr = \to [(2-Z)^3-1] \in (0,7)\to  \;,\; \gamma(\tcr) \equiv \gc = \frac{Z}{X^3} \go = 
   \frac{2\tsy(\go)}{3(t+\to)} \go < \go \;,\; \frac{\tsy(\gamma)}{\tad} = \frac{\gc}{\gamma}
\label{tgcr}
\end{equation}
For $Z < 1$, the electron cooling is SY-dominated until $\tcr$; at $\tcr$, the electron energy is
$\gc$ and the powers of the two cooling processes are equal; after $\tcr$, AD-cooling is dominant.
For $Z > 1$, when the electron cooling is AD-dominated at $t=0$, it can be shown using Equation (\ref{adsysol}) 
that $\tsy (\gamma) > \tad$ at any time, thus the electron cools adiabatically at all times. 

 Depending on which cooling process is dominant initially (SY-cooling if $Z < 1$, AD-cooling for $Z > 1$), the dominance 
at later times is established as following:
\begin{displaymath}
 Z < 1:  \quad  \left\{ \begin{array}{lllll}  
   \h [{\rm SY-cool}: \tsy(\gamma(t))<\tad(t)] & and \quad [\gamma(t)={\rm SY-sol}]           & for \quad [\gamma >\gc,\; t<\tcr,\;X<2-Z]  \\ 
   \h [{\rm AD-cool}: \tsy(\gamma)>\tad]       & and \quad [\gamma(t)={\rm \onethird SY-sol}] & for \quad [\gamma <\gc,\; t>\tcr,\;X>2-Z]  
  \end{array}  \right.            
\end{displaymath}
\begin{displaymath}
 \hspace*{-20mm} Z > 1:  \quad  \left\{ \begin{array}{lllll}  
   \h [{\rm AD-cool}: \tsy(\gamma)>\tad]       & and \quad [\gamma(t)={\rm AD-sol}]           & for \quad [t<\tt,\; X-1<Z/2] \\ 
   \h [{\rm AD-cool}: \tsy(\gamma)>\tad]       & and \quad [\gamma(t)={\rm \onethird SY-sol}] & for \quad [\tt<t,\; X-1>Z/2]  
  \end{array}  \right.            
\end{displaymath}

 Comparing these expectations with the expanded solution (Equations \ref{exp1}-\ref{exp3}) for electron cooling, 
we note that the condition for the SY-cooling solution to be asymptotically displayed at early times, $t \ll \to$,
is more restrictive than the condition for SY cooling to be dominant: $X < 2-Z$.
 Similarly, the conditions for the AD-cooling solution to be asymptotically manifested at early times: $t \ll \tt$ 
(or $X \ll Z/2+1$) and $Z > 1$, are more restrictive than the condition for AD-cooling to be dominant: $X > 2-Z$. 

 Furthermore, the expanded solution in Equations (\ref{exp1})-(\ref{exp3}) shows that a change in the evolution of the 
electron energy is not tied to the competition between the two cooling processes, which defines the critical electron 
energy $\gc$ (Equation \ref{tgcr}) that is crossed by the cooling electron at the critical time $\tcr$, but by the 
interplay between the SY and AD terms in the solution (Equation \ref{adsysol}) to the two-process cooling equation.
 Finally, at late times: $t \gg \max (\tt,\tcr)$, when AD-cooling is dominant ($X > 2-Z$ for $t > \tcr$), 
the solution to electron cooling is 1/3 of the SY-cooling solution ($X \gg Z/2+1$ for $t \gg \tt$) and not the 
AD-cooling solution.



\begin{references}
 \reference{} Axelsson M., Baldini L., Barbiellini G. et al, 2012, ApJ 757, L31 
 \reference{} Band D., 1997, ApJ 486, 928 
 \reference{} Bhat P., Fishman G., Meegan C. et al, 1994, ApJ 426, 604 
 \reference{} Crider A., Liang E., Smith I. et al, 1997, ApJ 479, L39 
 \reference{} Daigne F., Bosnjak Z., Dubus G., 2011, AA 526, 110 
 \reference{} Fenimore E., Zand J., Norris J., Bonnell J., Nemiroff R., 1995, ApJ 448, L101  
 \reference{} Ghirlanda G., Celotti A., Ghisellini G., 2003, A\&A 406, 879 
 \reference{} Granot J., 2016, ApJ 816, L20 
 \reference{} Guiriec S., Kouveliotou C., Daigne F. et al, 2015, ApJ 807, 148
 \reference{} Kaneko Y., Preece R., Briggs M. et al, 2006, ApJS 166, 298 
 \reference{} Kippen R., in't Zand J., Woods P. et al, 2004, AIP Conf. Proc. 727, 119 
 \reference{} Kumar P., Panaitescu A., 2000, ApJ 541, L51 
 \reference{} Kumar P., McMahon E., 2008, MNRAS 384, 33 
 \reference{} Lee A., Bloom E., Petrosian V., 2000, ApJS 131, 1 
 \reference{} Lloyd R., Petrosian V., 2000, ApJ 543, 722 
 \reference{} Medvedev M., Loeb A., 1999, ApJ 526, 706 
 \reference{} Medvedev M., 2000, ApJ 540, 704 
 \reference{} \Meszaros P., Rees M., 2000, ApJ 530, 292 
 \reference{} Nakar E., Ando S., Sari R., 2009, ApJ 703, 675 
 \reference{} Norris J., Nemiroff R., Bonnell J. et al, 1996, ApJ 459, 393 
 \reference{} Oganesyan G., Nava L., Ghirlanda G., Celotti A., 2017, ApJ 846, 137 
 \reference{} Panaitescu A., \Meszaros P., 2000, ApJ 544, L17 
 \reference{} Panaitescu A., 2019, ApJ 886, 106 (P19)
 \reference{} Panaitescu A., Vestrand W.T., 2022, ApJ, accepted (arXiv.org/abs/2209.11847)
 \reference{} Poolakkil S., Preece R., Fletcher C. et al, 2021, ApJ 913, 60 
 \reference{} Preece R., Briggs S., Mallozzi R. et al, 2000, ApJS 126, 19 
 \reference{} Ravasio M., Ghirlanda G., Nava L., Ghisellini G., 2019, A\&A 625, A60  
 \reference{} Ryde F., Axelsson M., Zhang B.B. et al, 2010, ApJ 709, L172 
 \reference{} Toffano M., Ghirlanda G., Nava L. et al, 2021, A\&A 652, A123
 \reference{} Zhang B., Huirong Y., 2011, ApJ 726, 90 
\end{references}
\end{document}